\documentclass{vldb}
\usepackage{balance}  
\usepackage[normalem]{ulem}
\usepackage{color}
\usepackage[ruled, vlined]{algorithm2e}
\usepackage{listings}
\usepackage{graphicx}

\DeclareGraphicsExtensions{.pdf,.jpeg,.png,.jpg}
\usepackage{subfig}

\usepackage{mdwlist}
\newenvironment{definition}[1][Definition.]{\begin{trivlist}
\item[\hskip \labelsep {\bfseries #1}]}{\end{trivlist}}

\usepackage{enumitem,calc}
\newcommand\preitem{\mdseries\textbullet\space}
\newlist{desclist}{description}{3}
\setlist[desclist,1]{format=\preitem\bfseries,leftmargin=\widthof{\preitem},style=sameline,parsep=1pt}
\newcommand\ipreitem{\mdseries}
\newlist{descdesp}{description}{3}
\setlist[descdesp,1]{format=\ipreitem\bfseries,leftmargin=0pt,style=sameline}
\newcommand{\REM}[1]{}

\begin{document}

\title{GRE: A Graph Runtime Engine for Large-Scale Distributed Graph-Parallel Applications}



%
%
%
%

\numberofauthors{1} 

\author{
%
%
\alignauthor
Jie Yan$\dagger~\ddagger$,	Guangming Tan$\dagger$,	Ninghui Sun$\dagger$\\
	\vspace{0.2cm}
       \affaddr{$\dagger$ Institute of Computing Technology, Chinese Academy of Sciences}\\
       \vspace{0.1cm}
       \affaddr{$\ddagger$ University of Chinese Academy of Sciences}
      	\email{\{yanjie, tgm, snh\}@ict.ac.cn}
\alignauthor
Guangming Tan\\
       \affaddr{Institute of Computing Technology, CAS}\\
       \email{tgm@ict.ac.cn}
\alignauthor
	Ninghui Sun\\
       \affaddr{Institute of Computing Technology, CAS}\\
       \email{snh@ict.ac.cn}
}
\date{30 July 2013}

\maketitle

\begin{abstract}
Large-scale distributed {\em graph-parallel} computing is challenging. On one hand, due to the irregular computation pattern and  lack of locality, it is hard to express parallelism efficiently. On the other hand, due to the scale-free nature, real-world graphs are hard to partition in balance with low cut. To address these challenges, several graph-parallel frameworks including Pregel and GraphLab (PowerGraph) have been developed recently. In this paper, we present an alternative framework, Graph Runtime Engine ({\tt GRE}). While retaining the vertex-centric programming model, {\tt GRE} proposes two new abstractions: 1) a {\tt Scatter-Combine} computation model based on active message to exploit massive fined-grained edge-level parallelism,  and  2) a {\tt Agent-Graph} data model based on vertex factorization to partition and represent directed graphs. {\tt GRE} is implemented on commercial off-the-shelf multi-core cluster. We experimentally evaluate {\tt GRE} with three benchmark programs (PageRank, Single Source Shortest Path and Connected Components) on real-world and synthetic graphs of millions$\sim$billion of vertices. Compared to PowerGraph, {\tt GRE} shows 2.5$\sim$17 times better performance on 8$\sim$16 machines (192 cores). Specifically, the PageRank in {\tt GRE} is the fastest when comparing to counterparts of other frameworks (PowerGraph, Spark,Twister) reported in public literatures. Besides, {\tt GRE} significantly optimizes memory usage so that it can process a large graph of 1 billion vertices and 17 billion edges on our cluster with totally 768GB memory, while PowerGraph can only process less than half of this graph scale.  
\end{abstract}

\section{Introduction}
Processing large-scale real-world graphs has become significantly important for mining valuable information and learning knowledge in many areas, such as data analytics, web search, and recommendation systems. The most frequently used algorithmic kernels, including path exploration (e.g. traversal, shortest paths computation) and topology-based iteration (e.g. page rank, clustering), are driven by graph structures. Parallelization of these algorithms is intrinsically different from traditional scientific computation that appeals to a data-parallel model, and has emerged as so-called {\em graph-parallel} \cite{powergraph} problems.

\begin{table*}[htp]
\center
\caption{Key Features of Distributed Graph-Parallel Systems}
\begin{tabular}{|c|c|c|c|}
\hline
Framework & Graph Data Model & Computation Model & Communication Model\\
\hline
Pregel & Directed, Hash-Mapping & Bulk Synchronous Parallel (BSP)& Message Passing\\
GraphLab & Undirected, Hash-Mapping & Asynchronous(Async.) & Distributed Shared Memory\\
PowerGraph & Undirected, Vertex-Cut & Gather-Apply-Scatter, BSP/Async. & Distributed Shared Memory\\
GRE & Directed, Agent-Graph & Scatter-Combine, BSP & Active Message\\
\hline
\end{tabular}
\label{tab_comp_all}
\vspace{-0.5cm}
\end{table*}%

In recent years we have witnessed an explosive growth of graph data. For example, the World Wide Web graph currently has at least 15 billion pages and one trillion URLs~\cite{webScale}. Also, the social network of Facebook~\cite{facebookScale} has over 700 million users and 140 billion social links. Even to store only the topology of such a graph, the volume is beyond TeraBytes (TB), let alone rich metadata on vertices and edges. The efficient processing of these graphs, even with linear algorithmic complexity, has scaled out capacity of any single commodity machine. Thus, it is not surprising that distributed computing has been a popular solution to graph-parallel problems~\cite{pregel, graphlab, powergraph, spark}. However, since the scale-free nature of real-world graphs, we are facing the following two major challenges to develop high performance graph-parallel algorithms on distributed memory systems \cite{challenges}.
\begin{desclist}
\item {\bf Parallelism expressing.} {\it Graph-parallel algorithms often exhibit random data access, very little work per vertex and a changing degree of parallelism over the course of execution\cite{pregel}, making it hard to express parallelism efficiently}. Note the facts that graph-parallel computation is data-driven or dictated by the graph topology, and real-world graphs are unstructured and highly irregular (known as scale-free, e.g. low-diameter, power-law degree distribution). Thus, on one hand, from the view of programming, graph-parallel computation can't fit well in traditional parallelization methods based on decomposing either computational structure or data structure. For example, MapReduce~\cite{mapreduce}, the widely-used data-parallel model, can't efficiently process graphs due to the lack of support to random data access and iterative execution. On the other hand, from the view of performance, the lack of locality makes graph-parallel procedures memory-bound on the shared memory system and network-bound on the distributed system. Meanwhile, current computer architecture is evolving to deeper memory hierarchy and more processing cores, seeking more locality and parallelism in programs. Considering both ease of programming and affinity to the system architecture, it is challenging to express parallelism efficiently for graph-parallel algorithms.
\item {\bf Graph data layout.} {\it Real-world graphs are hard to partition and represent in a distributed data model.} The difficulty is primarily from graph's scale-free nature, especially the power-law degree distribution. The power-law property implies that a small fraction of vertices connect most of edges. For example, in a typical power-law graph with degree distribution $P(d) \propto d^{-\alpha}$ where $\alpha=2$, 20\% of vertices connect more than 80\% of edges. As identified by previous work~\cite{hardPar1, hardPar2}, this skew of edge distribution makes a balanced partitioning with low edge-cut difficult and often impossible for large-scale graphs in practice. Reference~\cite{streamingPar} thoroughly investigated streaming partitioning methods with all popular heuristics on various datasets. However, their released results show that for scale-free graphs the edge-cut rate is very high, only slightly lower than a random hashing method. As a consequence, graph distribution leads to high communication overhead and memory consumption.
\end{desclist}

To address the above challenges, several distributed graph-parallel frameworks have been developed. Basically, a framework provides a specific compution abstraction with a functional API to express graph-parallel algorithms, leaving details of parallelization and data transmission to the underlying runtime system. Representative frameworks include Pregel\cite{pregel}, GraphLab\cite{graphlab}, PowerGraph\cite{powergraph} and matrix-based packages~\cite{pegasus}\cite{kdt}. These efforts target three aspects of graph-parallel applications: {\em graph data model, computation model} and {\em communication model}. Table~\ref{tab_comp_all} summarizes key technical features of the three frameworks ({\tt Pregel}, {\tt GraphLab}, {\tt PowerGraph}), as well as {\tt GRE} to be presented in this paper. 

Pregel~\cite{pregel} acts as a milestone of graph-parallel computing. It firstly introduced the {\em vertex-centric} approach that has been commonly adopted by most of later counterparts~\cite{signal_collect, graphlab, powergraph, x-stream} including ours. This idea is so-called {\em think like a vertex} philosophy, where all active vertices perform computation independently and interact with their neighbors typically along edges. Pregel organizes high-level computation in Bulk Synchronous Parallel (BSP) super-steps and adopts message passing model for inter-vertex communication. GraphLab supports asynchronous execution any more, but uses distributed shared memory such that the vertex can directly operates its edges and neighbors. For both Pregel and GraphLab, the vertex computation follows a common pattern where an active vertex 1) collects data from its in-edges, 2) updates its states, 3) puts data on its out-edges and signals the downwind neighbors. Moreover, PowerGraph, the descent of GraphLab, summarized the above vertex procedure as a Gather-Apply-Scatter ({\tt GAS}) paradigm, and explicitly decomposes it into three split phases, which exposes potential edge-level parallelism. However, the {\tt GAS} abstraction inherently handles each edge in a two-sided way that requires two vertices' involvement (Scatter and Gather respectively), which leads to intermediate data storage and extra operations. Instead, {\tt GRE} proposes a new {\tt Scatter-Combine} computation model, where the previous two-sided edge operations are reduced to one active message\cite{active_messages}. Besides, compared to pure message passing or distributed shared memory model, active message has better affinity to both local multi-core processors and remote network communication.

With respect to distributed graph data model, most of previous frameworks including Pregel and GraphLab use a simple hash-mapping strategy where each vertex and its edges are evenly assigned to a random partition. While being fast to load and distribute graph data, this method leads to a significantly high edge-cut rate. Recently, PowerGraph introduces {\em vertex-cut} in which vertex rather than edge spans multiple partitions. {\em Vertex-cut} can partition and represent scale-free graphs with significantly less resulting communication. However, it requires to maintain strict data consistency between master vertex and its mirrors. Instead, {\tt GRE} proposes a new {\tt Agent-Graph} model, where data on {\tt agent} is temporal so that the consistency issue is avoided.

In a summary, {\tt GRE} inherits the {\em vertex-centric} programming model, and specifically makes the following major contributions:
\vspace{-0.15cm}
\begin{itemize*}
\item {\tt Scatter-Combine}, a new graph-parallel computation model. It is based on active message and fine-grained edge-level parallelism. (Sec. \ref{sec:comp_model})
\item {\tt Agent-Graph}, a novel distributed directed graph model. It extends the original graph with {\em agent} vertices. Specifically, it has no more and typically much less communication than PowerGraph's {\em vertex-cut}. (Sec.~\ref{sec:graph_model})
\item Implementation of an efficient runtime system for {\tt GRE} abstractions. It incorporates several key components of data storage, one-sided communication and fine-grained data synchronization. (Sec.~\ref{sec:runtime})
\item A comprehensive evaluation of three benchmark programs (PageRank, SSSP, CC) and graph partitioning on real-world graphs, demonstrating {\tt GRE}'s excellent performance and scalability. Compared to PowerGraph, {\tt GRE}'s performance is 2.5$\sim$17.0 times better on 8$\sim$16 machines. Specifically, {\tt GRE}'s PageRank takes $2.19$ seconds per iteration on 192 cores (16 machines) while PowerGraph reported $3.6$ seconds on 512 cores (64 machines)\cite{powergraph}. {\tt GRE} can process a large graph of one billion vertices on our machine with 768GB memory while PowerGraph cannot make it. (Sec.~\ref{sec:evaluation})
\end{itemize*}
\vspace{-0.20cm}
\section{Background and Motivation}
In this section, we formalize the procedure of vertex-centric graph computation and then present motivation of this work.
\subsection{Preliminaries}
\label{modeling_graph_com}
Graph topology can be represented as $G = (V, E)$, where $V$ is the set of vertices and $E$ the set of edges. Associating $G$ with metadata on vertices and edges, we have a property graph $G' = (V, E, V.P, E.P)$, where $V.P$ and $E.P$ are properties of vertices and edges respectively. Property graph is powerful enough to support all known graph algorithms. In this paper, all edges are considered as directed (one undirected edge can be transformed into two directed edges). For simplicity, we define the following operations:
\vspace{-0.20cm}
\begin{itemize*}
    \item $e_{out}:V \rightarrow E$: return the set of a vertex's out-edges;
    \item $e_{in}~:V \rightarrow E$: return the set of a vertex's in-edges;
    \item $v_{out}:E \rightarrow V$: return the source vertex of an edge;
    \item $v_{in}~:E \rightarrow V$: return the target vertex of an edge;
    \item $\epsilon~~~:E \cup V \times R \rightarrow E \cup V$: filter a set of vertices or edges with rule $R$, and return a subset.
\end{itemize*}

As an abstraction, the computation $\mathcal{F}$ on some vertex $v$ can be described as:
\begin{equation}\label{e1}
    \begin{split}
    v.state = \mathcal{F}(v.state,\ \{e_i.state\},\ \{v_i.state\})\ ,\\
    where\ e_i \in e_{in}(v)\ and\ v_i \in v_{out}\circ e_{in}(v).
    \end{split}
\end{equation}
For example, an instance of PageRank~\cite{pagerank} computation on $v$ can be described as:
\begin{equation}\label{pr-1}
v.pr = 0.15 + 0.85*(\sum_{v_i} v_i.pr / |e_{out}(v_i)|).
\end{equation}
%

In the vertex-centric approach, calculation of Equation.~\ref{e1} is encoded as so-called {\em vertex-program}. Each vertex executes its {\em vertex-program} and can communicate with its neighbors. Depending on pre-defined synchronization policies, active vertices are scheduled to run in parallel. A common pattern followed by {\em vertex-program} is Gather-Apply-Scatter ({\tt GAS})\cite{powergraph} (or Signal/Collect\cite{signal_collect}). It translates Equation.~\ref{e1} to the following three conceptual steps \footnote{Strictly speaking, {\tt GAS} is not necessarily equivalent to Equation.~\ref{e1}, as {\tt GAS} requires that the vertex computation can be factored into a generalized sum of products.}:
\begin{itemize*}
\item {\bf G}ather. Vertex $v$ collects information from in-edges and upwind neighbor vertices by a generalized sum $\bigoplus$, resulting in $v.sum$:
\begin{equation}\label{gather}
	\begin{split}
		v.sum \leftarrow \bigoplus_{\substack{
		e_i \in e_{in}(v)\\
		v_i = v_{out}(e_i)
		} }g(v.state, e_i.state, v_i.state).
	\end{split}
\end{equation}
\item {\bf A}pply. Vertex $v$ recomputes and updates its state:
\begin{equation}\label{apply}
	\begin{split}
	v.state \leftarrow a(v.state, v.sum).
	\end{split}
\end{equation}
\item {\bf S}catter. Vertex $v$ uses its newest state to update the out-edges' states and signals its downwind neighbors:
\begin{equation}\label{scatter}
	\begin{split}
	\forall e'_i \in &e_{out}(v),~v'_i = v_{in}(e'_i) :\\
	&e'_i.state \leftarrow s(v.state, e'_i.state, v'_i.state).
	\end{split}
\end{equation}
\end{itemize*}

In steps of gather and scatter, the vertex $v$ communicates with its upwind and downwind neighbors respectively. Conventionally, there are two methods to do the communication, that message passing (Pregel~\cite{pregel} and other similar systems like Spark~\cite{spark}, Trinity~\cite{trinity}) and distributed shared memory (GraphLab~\cite{graphlab} and its descendant PowerGraph~\cite{powergraph}). In distributed shared memory, remote vertices are replicated in local machine as {\em ghost}s or {\em mirror}s, and data consistency among multiple replications is maintained implicitly.

\subsection{Motivation}
\label{sub:motivation}
Although {\tt GAS} model provides a clear abstraction to reason about parallel program semantics, it may sacrifice storage and performance. Note that {\tt GAS} conceptually split the information transferring on an edge into two phases executed by two vertices, i.e. the source vertex changes the edge state in its scatter phase and then the target vertex reads the changed state in its gather phase. The above asynchrony of operations on the edge requires storage of intermediate edge states and leads to extra operations that hurt performance. In Pregel this happens across two super-steps and leads to the large storage of intermediate messages, while in GraphLab it requires not only out-edge storage but also redundant in-edge storage and polling on all in-edges. 

However, we identify that in a message model the separation of gather and scatter is not necessary, as long as the operator $\bigoplus$ in gather is commutative. To illustrate this point, again we take PageRank as an example and rewrite its vertex computation in Equation.~\ref{pr-1} as follow:
\begin{equation}
	\begin{split} \label{pr-2}
		&(a): msg_i \leftarrow v_i.pr / |e_{out}(v_i)|,\\
		&(b): v.sum \leftarrow \sum_{\substack{
		e_i \in e_{in}(v)
		} } msg_i, \\
		&(b): v.pr \leftarrow 0.15+0.85*v.sum.
	\end{split}
\end{equation}
Equation.~\ref{pr-2}a is the message sent to $v$ by vertex $v_i$. In Equation.~\ref{pr-2}b, the $\sum$ operation is commutative, which means the order of computing $msg_i$ does no matter. Once a message comes, it can be immediately computed. In practice, given the fact that {\em graph-parallel computation is essentially driven by data flow on edges}, the $\bigoplus$ is naturally commutative.

\begin{figure*}[htbp]
\centering
\includegraphics[scale=0.45]{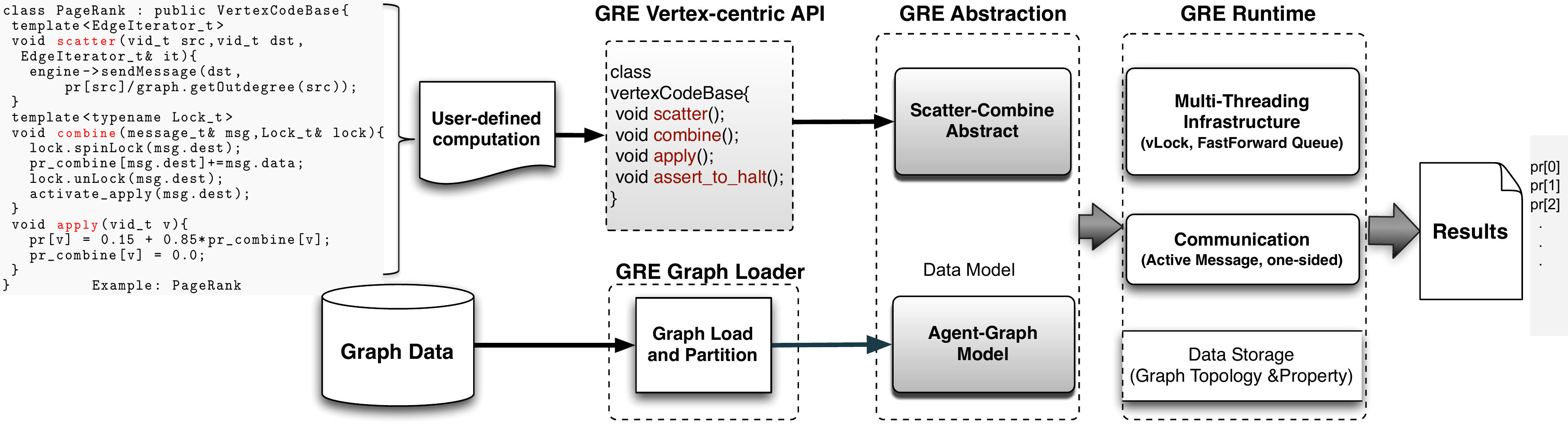}
\caption{Graph Runtime Engine (GRE) Architecture.}
\vskip -6pt
\label{gre-arch}
\end{figure*}

From the above analysis, we can induce a dataflow execution model that leverages active message approach~\cite{active_messages}. An active message is a message containing both data and encoded operations. Now the gather phase then can be broken into a series of discrete asynchronous active messages. An active message can be scheduled to run once it reaches the destination vertex. Specifically, for a message on the edge $e_{u \rightarrow v}$, when $u$ and $v$ are owned by the same machine, the message can be computed in-place during $u$'s scatter phase. Note that multiple active messages may operate on the same destination vertex simultaneously, leading to data race. As detailed in later sections, we shall handle this issue with a vertex-grained lock mechanism.

Basically, the dataflow model has two significant advantages. First, it transforms the two-sided communication to one-sided, bypassing the intermediate message storage and signaling. This optimization dramatically reduces the overhead in both shared memory and distributed environment. Second, it enables fine-grained edge-level parallelism that takes advantage of multi-core architecture.

Therefore, we propose {\tt Scatter-Combine}\footnote{The naming of {\tt Scatter-Combine} implies an active message on an edge, where {\tt Scatter} generates a message by source vertex and automatically incurs a {\tt Combine} on target vertex.} (Sec.\ref{sec:comp_model}), an alternative graph-parallel abstraction that is more performance-aware. Besides, we notice that in Equation.~\ref{pr-2}b, the $\sum$ is also associative. In fact, for most graph-parallel problems, not only commutativity but associativity is also satisfied by the generalized sum $\bigoplus$ in Equation.~\ref{gather}. This fact has been widely realized as the basis of Pregel's message {\em combiner} and PowerGraph's {\em vertex-cut} mechanism. Based on the associativity and commutativity of $\bigoplus$, we develop {\tt Agent-Graph} (Sec.\ref{sec:graph_model}), a novel distributed graph data model that can effectively partition and represent scale-free graphs in the distributed environment. {\tt Agent-Graph} is closely coupled with active message approach. This evolution motivates the development of  {\bf G}raph {\bf R}untime {\bf E}ngine ({\tt GRE}).

\section{Overview of GRE Architecture}
\label{sec:overview}
In this section, we give an overview of {\tt GRE} system, as shown in Fig.~\ref{gre-arch}. {\tt GRE} is implemented in C++ templates. It consists of graph loader, abstraction layer and underlying runtime layer. The essentials of {\tt GRE} are the {\tt Scatter-Combine} computation model and distributed {\tt Agent-Graph} data model.

As noted in the previous sections, {\tt GRE} inherits the {\em vertex-centric} programming model. The programming interface is a set of simple but powerful functional primitives. To define a graph-parallel algorithm, users only need to define the vertex computation with these primitives. In Fig.~\ref{gre-arch}, user-defined PageRank program is presented as an example, which is as simple as its mathematic form in Equation.~\ref{pr-2}.  The user-defined program, as template parameters, is then integrated into {\tt GRE} framework and linked with runtime layer.

Internally, {\tt GRE} adopts distributed {\tt agent-graph} to represent graph topology, and column-oriented storage~\cite{cos} to store vertex/edge property. The graph loader component loads and partitions graphs into internal representation. 

The runtime layer provides infrastructure supporting {\tt GRE}'s abstractions. With thread pool (or thread groups) and fine-grained data synchronization, {\tt GRE} can effectively exploit the massive edge-level parallelism expressed in {\tt Scatter-Combine} model. Besides, {\tt GRE} adopts one-sided communication to support active message and override the network communication overhead with useful computation.

\section{Computation Model}
\label{sec:comp_model}
{\tt GRE} models the graph-parallel computation in {\tt Scatter- Combine} abstraction. As noted in section~\ref{sub:motivation}, it realizes the fact that for a broad set of graph algorithms, vertex-centric computation can be factored into independent edge-level functions, and thus transforms the bulk of vertex computation into a series of active messages. 

\subsection{Scatter-Combine Abstraction}
In {\tt Scatter-Combine}, each active vertex is scheduled to compute independently and interacts with others by active messages. Like in Pregel and GraphLab, the major work of {\tt GRE} programming is to define vertex-computation. 

{\tt Scatter-Combine} provides four primitives: {\em scatter}, {\em combine}, {\em apply} and {\em assert\_to\_halt}. Their abstract semantics are specified in Alg.~\ref{alg:primitives}, and run in the context of Alg.~\ref{alg:logics}. To implement a specific algorithm, users only need to instantiate these primitives. Each vertex alternately carries computation following the logics in Alg.~\ref{alg:logics}, where the procedure is divided into two phases, {\tt scatter-combine} and {\tt apply}. Each vertex implicitly maintains two state variables, one for {\tt scatter} and the other for {\tt apply}, to decide whether to participate in the computation of the relevant phase.

During {\tt scatter-combine} phase, a vertex, if being active to scatter, modifies its outgoing edges's states and scatters data to its downwind neighbors by active messages. As emphasized before, active message is the essential of {\tt Scatter-Combine} computation. It is edge-grained and defined by primitives {\em scatter} and {\em combine}. As shown in Alg.~\ref{alg:primitives}, the {\em scatter} primitive generates an active message that carries a {\em combine} operation, while the {\em combine} primitive defines how the message operates on its destination vertex. Besides, {\em combine} is able to activate the destination vertex for a future {\em apply}. Note that unlike previous message passing in Pregel, active message is one-sided. That is, when the vertex $u$ sends a message to $v$, it directly operates on $v$ without $v$'s involvement. Conceptually, the {\em scatter} doesn't necessarily wait its paired {\em combine} to return. An active vertex may execute {\em scatter} on all or a subset of its out-edges. After finishes all desired {\em scatter} operations, the vertex calls user-defined {\em assert\_to\_halt} to deactivates itself optionally. Generally, in traversal-based algorithms {\em assert\_to\_halt} is defined to deactivate for scatter, and for iterative algorithms such as PageRank it is defined to keep the vertex active.
\vspace{-0.25cm}
\IncMargin{-0.6em}
\begin{algorithm}[h]
\caption{Specification of Scatter-Combine Primitives\label{alg:primitives}}
\IncMargin{0.6em}
\small{
\SetAlgoVlined
	struct $message$~\{\\
		\quad$dest$ \tcp*[l]{destination vertex}
		\quad$data$ \tcp*[l]{message data}
	\}\;
	{\bf scatter} ($u, v, e_{u \rightarrow v}$):\\
		\quad$message:msg$\;
		\quad$msg.dest \leftarrow v$\;
		\quad$msg.data \leftarrow s(u.state,~e_{u \rightarrow v}.state)$\;
		\quad {\tt send}($msg$, {\bf combine})\tcp*[l]{active message}
	{\bf combine} ($message: msg$):\\
		\quad$v \leftarrow msg.dest$\;
		\quad$v.sum \leftarrow v.sum \bigoplus c(v.state, msg.data)$\;
		\quad$activate\_apply$($v$); \tcp{optional}
	{\bf apply} ($v$):\\
		\quad$v.state \leftarrow a(v.state, v.sum)$\;
		\quad$activate\_scatter$($v$); \tcp{optional}
	{\bf assert\_to\_halt} ($v$):\\
		\quad$deactivate\_scatter$($v$); \tcp{optional}
}
\end{algorithm}
\vspace{-0.6cm}
\begin{algorithm}[h]
\caption{Logics of Vertex-Computation \label{alg:logics}}
\small{
\SetAlgoVlined
\SetKwInOut{Input}{Input}
	\Input{centric vertex $u$}
	\Begin{
	\tcp{Phase 1: Scatter-Combine}
	\If{{\tt active\_for\_scatter}(u)}
	{	
		\ForEach{out-edge $e_{u \rightarrow v}$}
	 	{
			{\color{red}$scatter(u,~v,~e_{u \rightarrow v})$}\;	
		}
		{\color{red}$assert\_to\_halt(u)$}\;
	}
	\tcp{Phase 2: Apply}
	\If {{\tt active\_for\_apply}(u)}
	{		
		{\color{red}$apply(u)$}\;
		{\tt deactivate\_apply($u$)}\;
	}
	}
}
\end{algorithm}

During {\tt apply} phase, a vertex, if being active to apply, executes an {\em apply} and then sets its apply-phase state inactive. In the {\em apply} procedure, the vertex recomputes its state ({\em v.state}) with intermediate results ({\em v.sum}) accumulated in the previous phase, and optionally activates itself to participate in the {\tt scatter-combine} phase of next round.

{\tt GRE} adopts bulk synchronous parallel execution. Like Pregel, {\tt GRE} divides the whole computation into a sequence of conceptual super-steps. In each super-step, {\tt GRE} executes the above two phases in order. During each phase, all active vertices run in parallel.

The computation is launched by initializing any subset of vertices as source that are activated for {\tt scatter}. During the course of whole execution, more vertices are either activated to compute or deactivated. At the end of a super-step, if no vertex is active for further {\tt scatter-combine}, the whole computation terminates.

\begin{figure}[h]
	\centering	
	\subfloat[GAS]{
		\includegraphics[scale=0.28]{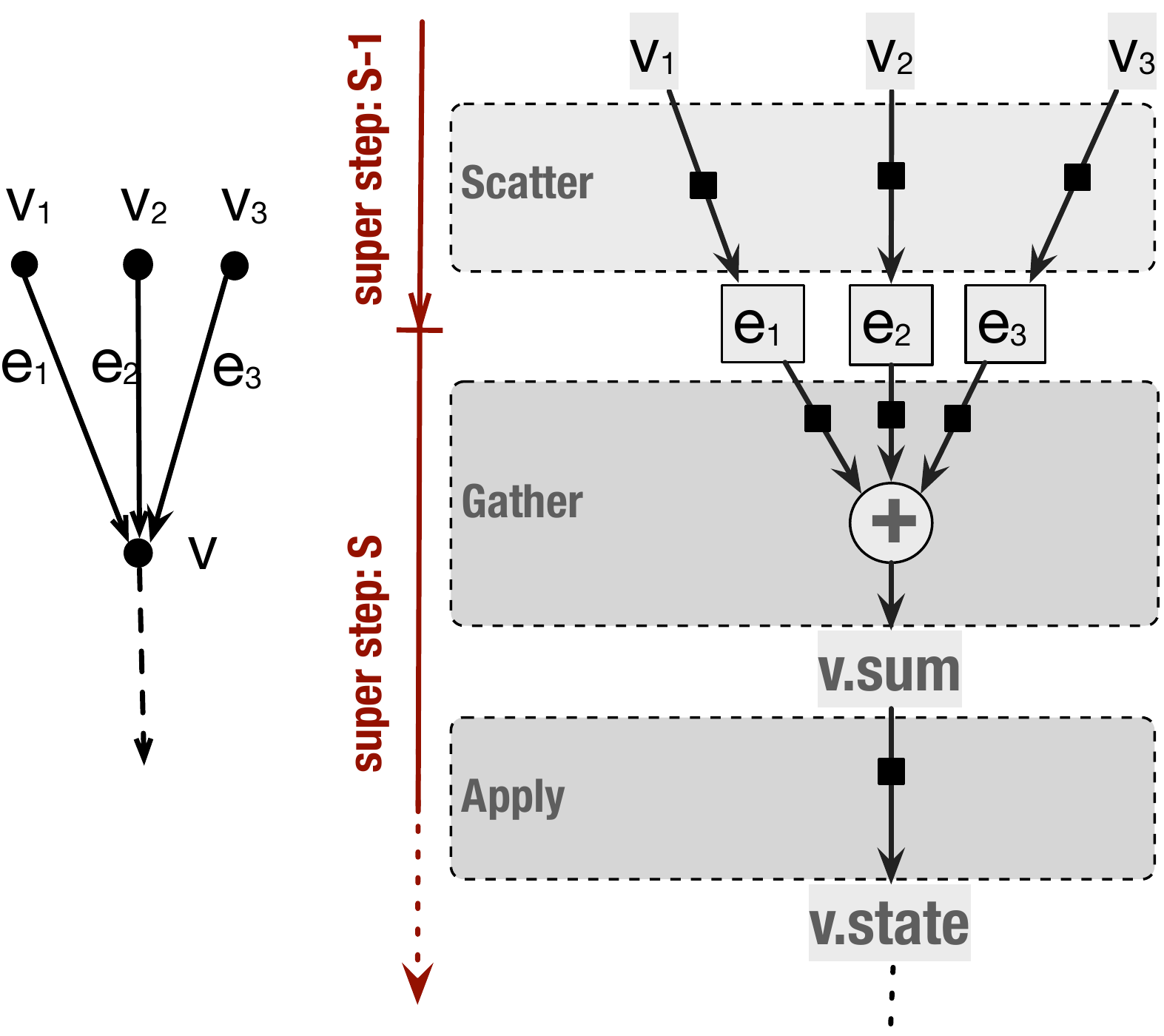}
		\label{subfig:pr-code}
	}
	\hspace{0.2cm}
	\subfloat[Scatter-Combine]{
		\includegraphics[scale=0.28]{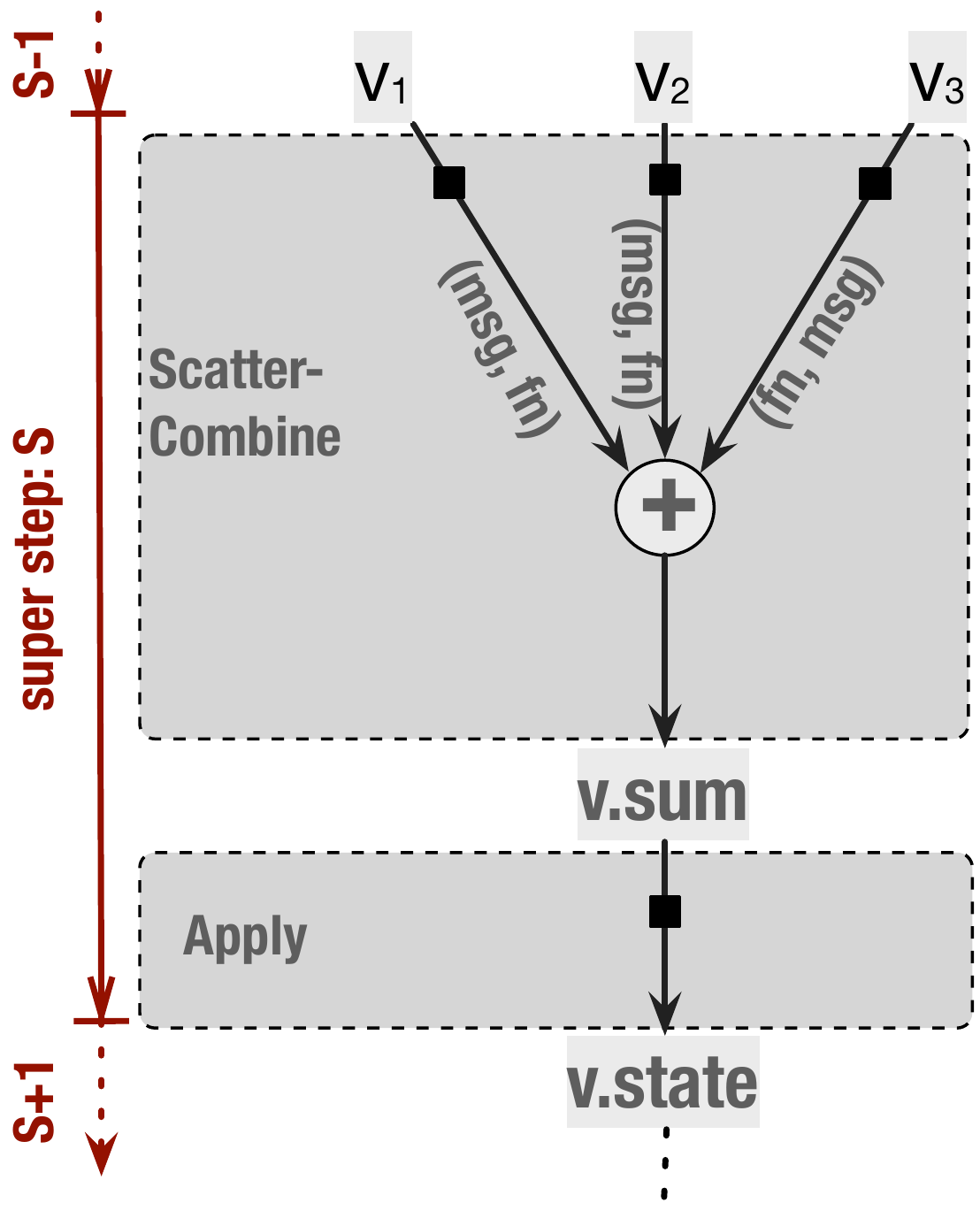}
		\label{subfig:sssp-code}
	}
\caption{Comparison of GAS and Scatter-Combine Models. This example illustrates how a vertex $v$ of a directed graph is computed in BSP mode.}
\label{fig:compare-models}
\end{figure}

Note that there are essential differences between {\tt GAS} and {\tt Scatter-Combine}. Fig.~\ref{fig:compare-models} illustrates their execution flow on an vertex $v$ in bulk synchronous parallel execution. We assume $v$'s state is computed in super step $S$. In {\tt GAS} model, at super-step $S$-$1$, the upwind neighbors ($v_1$, $v_2$ and $v_3$) have executed {\em scatter} and put data on $v$'s in-edges ($e_1$, $e_2$ and $e_3$). At super-step $S$ when $v$ is active, it {\em gather}s data by polling its in-edges and accumulates them in a local variable ($v.sum$ here). We can see that processing an edge needs a pair of {\em scatter} and {\em gather} executed by two vertices respectively, which crosses two super-steps and requires storage of all intermediate data on edges. As a significant progress, in {\tt Scatter-Combine}, the operation of {\em gather} is encoded in an active message that can automatically execute without the target vertex's involvement, namely {\em combine}. In this example, during {\tt scatter-combine} phase of super-step $S$, $v$'s upwind neighbors {\em scatter} active messages that directly execute {\em combine} on $v$, and finally $v$ simply updates itself by an {\em apply} during the {\tt apply} phase. 
\vspace{-0.25cm}
\vspace{-0.5cm}

\subsection{Applications}
\begin{figure}[t]
	\centering	
	\subfloat{
		\includegraphics[scale=0.65]{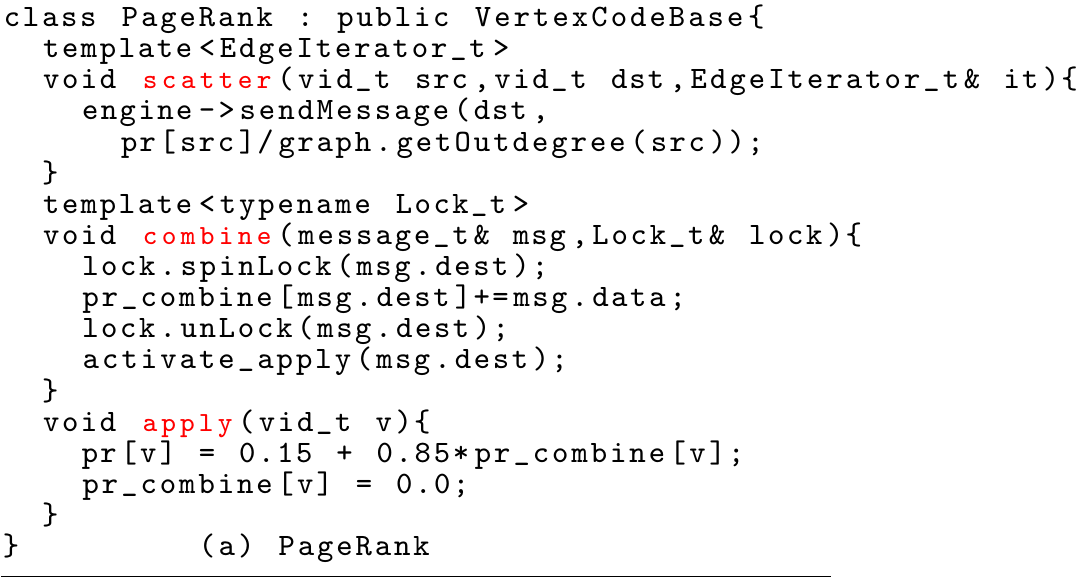}
		\label{subfig:pr-code}
	}
	\vspace{-0.2cm}
	\\
	\subfloat{
		\includegraphics[scale=0.65]{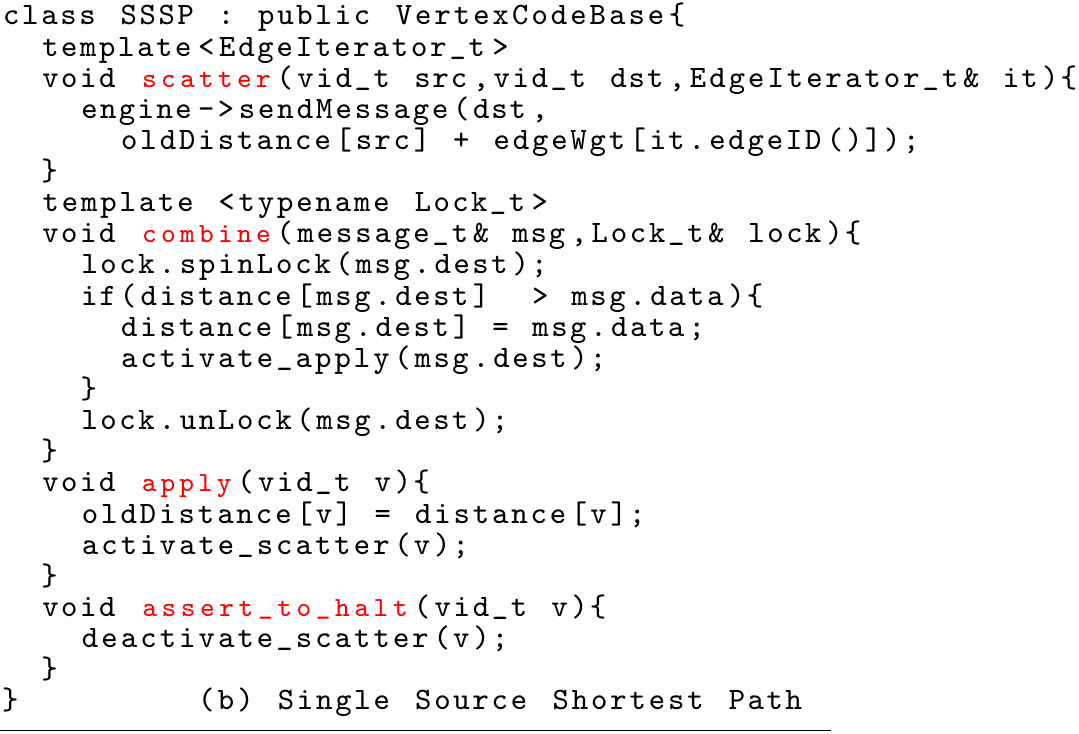}
		\label{subfig:sssp-code}
	}
	\vspace{-0.2cm}
	\\
	\subfloat{
		\includegraphics[scale=0.65]{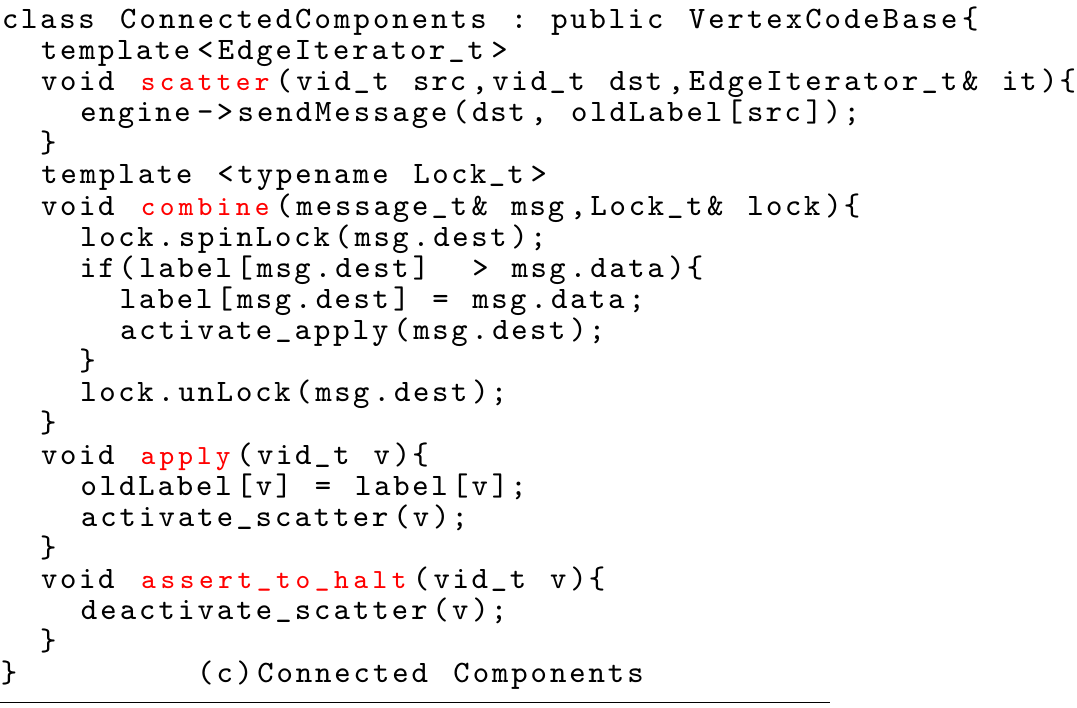}
		\label{subfig:cc-code}
	}
\caption{C++ codes of PageRank, SSSP and Connected Components in Scatter-Combine model.}
\vspace{-0.60cm}
\label{fig:applications}
\end{figure}

Programming with {\tt Scatter-Combine} model is very convenient. For instance, to implement PageRank, we directly translate the formulas \ref{pr-2}-a, \ref{pr-2}-b and \ref{pr-2}-c in Equation.~\ref{pr-2} into primitives {\em scatter}, {\em combine}, and {\em apply}, as shown in Fig.\ref{subfig:pr-code}. Besides, this figure presents implementations of other two algorithms that we use as benchmark in later evaluation.
\begin{descdesp}
\item[{\em Single Source Shortest Path~\cite{bellman1956routing, sssp}.}] {\tt GRE}'s SSSP implementation, given in Fig.~\ref{subfig:sssp-code}, is a variant of Bellman Ford label correcting algorithm. It is a procedure of traversal, starting from a given source vertex, visiting its neighbors and then neighbors's neighbors in a breadth first style, and continuing until no vertices change their states. When a vertex is visited, if its stored distance to source is larger than that of the new path, its path information is updated.
\item[{\em Connected Components~\cite{introAlg}.}] {\tt GRE} implements Connected Components as an example of label propagation. Fig.~\ref{subfig:cc-code} shows the Connected Components on undirected graphs. For each connected component, it is labeled by the smallest ID of its vertices. In the beginning, each vertex is initialized as a component labeled with its own vertex ID. The procedure then iteratively traverses the graph and combines new found connected components. Its algorithmic procedure is similar to SSSP,  but initiates all vertices as sources and typically converges in fewer number of iterations.
\end{descdesp}

{\tt Scatter-Combine} model can express most graph-parallel computation efficiently, including traversal-based path exploration and most iterative algorithms. In fact, since graph-parallel computation is internally driven by data flow on edges~\cite{challenges}, edge-parameterized vertex factorization is widely satisfied. We found that all examples of Pregel in \cite{pregel} satisfy vertex factorization, such as Bipartite Matching~\cite{bipartite-match} and Semi-clustering~\cite{pregel}, thus can be directly implemented in {\tt GRE}. For some graph algorithms that contain other computation except for graph-parallel procedure, extension to basic {\tt Scatter-Combine} model is required. For example, with simple extension of backward traversal on transposed graphs, {\tt GRE} implements multi-staged algorithms like Betweenness Centrality~\cite{bc} and Strong Connected Components.

Besides, with technologies that were proposed in reference \cite{optimize_pregel} to complement vertex-centric parallelism in Pregel-like systems, {\tt GRE} is able to efficiently implement algorithms including Minimum Spanning Forests, Graph Coloring and Approximate Maximum Weight Matching. However, like Pregel, {\tt GRE} only supports BSP execution, and thus can't express some asynchronous algorithms in PowerGraph. 

\section{Distributed Graph Model}
\label{sec:graph_model}
In this section we propose the distributed {\tt Agent-Graph} model that extends original directed graphs with $agent$ vertices. {\tt Agent-graph} is coupled with the message model, and is able to efficiently partition and represent scale-free graphs.

\subsection{Agent-Graph Model}
\label{sub:agent-graph-model}
There is a consensus that the difficulty of partitioning a real-world graph comes from its scale-free property. For big-vertex whose either in-degree or out-degree is high, amounts of vertices in remote machines send messages to it or it sends messages to amounts of remote vertices. Lack of low-cut graph partition, these messages lead to heavy network communication and degrades performance significantly.

\begin{figure}[h]
	\centering	
	\subfloat[Combiner ($v'$)]{
		\includegraphics[scale=0.35]{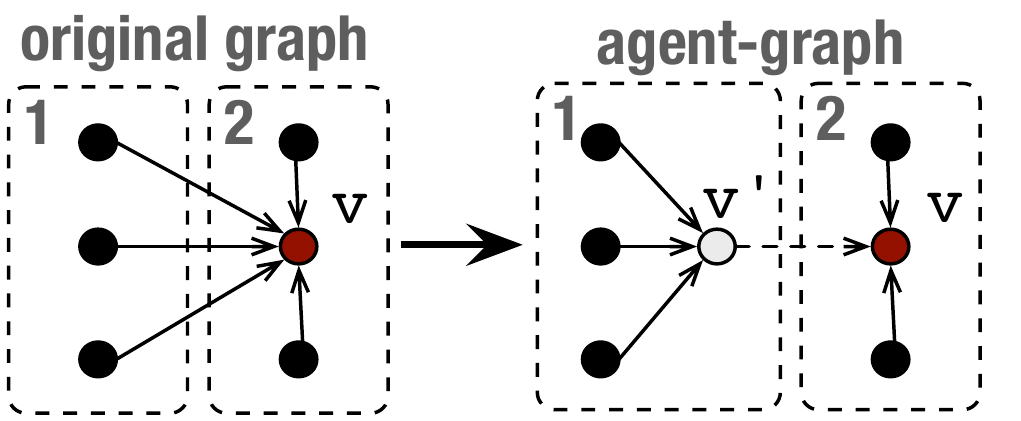}
		\label{subfig_combiner}
	}
	\hspace{0.75em}
	\subfloat[Scatter ($v'$)]{
		\includegraphics[scale=0.35]{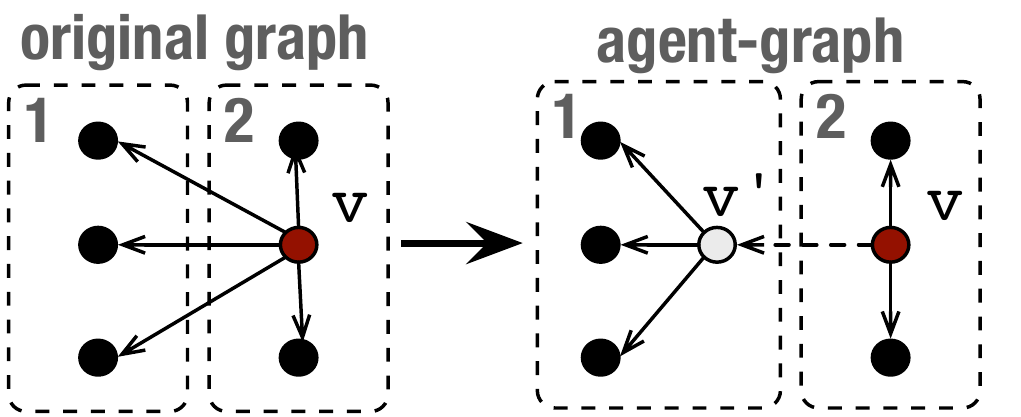}
		\label{subfig_scatter}
	}
\caption{Agent Model.}
\label{fig:agent}
\vspace{-0.25cm}
\end{figure}

To crack the big-vertex problem, {\tt GRE} introduces a new strategy -- {\em agent}. The basic idea is demonstrated in Fig.~\ref{fig:agent}. There are two kinds of agents, i.e. {\em combiner} and {\em scatter}. In the given examples, there are two machines (or graph partitions) where $v$ is the big-vertex and owned by machine 2. In Fig.~\ref{subfig_combiner}, $v$ is a high in-degree vertex, and many vertices in machine 1 send messages to it. By introducing a {\em combiner} $v'$, now messages previously sent to $v$ are first {\em combined} on $v'$ and later $v'$ sends a message to $v$. In this example, the {\em combiner} reduces network communication cost from three messages to one. Similarly, in Fig.~\ref{subfig_scatter}, $v$ no longer directly sends messages to remote vertices in machine 1 but only one message to a {\em scatter} agent $v'$ who then delivers messages to vertices in machine 1. Also, the {\em scatter} agent reduces messages on network from three to one.

Based on the idea of {\em agent}, we propose {\tt Agent-Graph}, simply denoted as $G^{\alpha}$. $G^{\alpha}$ treats {\em agent}s as special vertices and extends the original graph topology  $G$. For simplicity, we call the vertex in original graph as {\em master} vertex. Each {\em master} vertex is uniquely owned by one partition but can have arbitrary {\em agent}s in any other partitions. Each {\em agent} has an directed edge connected with its {\em master}. One thing to note is that the term of {\em agent} is completely transparent to programmers, and only makes sense to {\tt GRE}'s runtime system.

Now we give the formal description of $G^\alpha$. We assume graph $G$ has been divided into $k$ parts, say $\{G_1, G_2, ..., G_k\}$.
\begin{definition}
In $G_i$, for any set of directed edges pointing to $v \in V$, if $v$ is not owned by $G_i$, we set an agent $v_c$ for $v$ and do the following transformation: let the edges redirect to $v_c$, and add a directed edge $e_{v_c \rightarrow v}$. Then $v_c$ is a {\tt combiner} of $v$. A combiner may have arbitrary in-edges but only one out-edge that points to its {\em master} vertex.
\end{definition}
\begin{definition}
In $G_i$, for any set of directed edges that start from $v \in V$ and point to a set of vertices owned by another partition $G_j$, we set an agent $v_s$ on remote $G_j$, and do the following transformation: move these edges to $G_j$, and add a directed edge $e_{v \rightarrow v_s}$ on $G_i$. Then $v_s$ is a {\tt scatter} of $v$. A {\em scatter} may have arbitrary out-edges but only one in-edge that comes from its {\em master} vertex.
\end{definition}
\begin{definition}
Let $V_s$ be the set of {\it scatter}s and $V_c$ the set of {\em combiner}s, an {\tt agent-graph} $G^{\alpha} \triangleq \{V^{\alpha}, E^{\alpha}\}$, where $V^{\alpha} = V \bigcup V_c \bigcup V_s$ and $E^{\alpha} = E \bigcup \{(v_c, v) | v_c \in V_c , v \in V\} \bigcup \{(v, v_s) \\| v \in V, v_s \in V_s\} \bigcup \{(v_s, v_c) | v_s \in V_s, v_c \in V_c\}$. Note that according to the definitions of {\em scatter} and {\em combiner}, an edge from $v_s$ to $v_c$ is allowed, but there never exist an edge from $v_c$ to $v_s$.
\end{definition}

\begin{figure}[h]
	\centering	
	\subfloat[Agent-graph]{
		\includegraphics[scale=0.28]{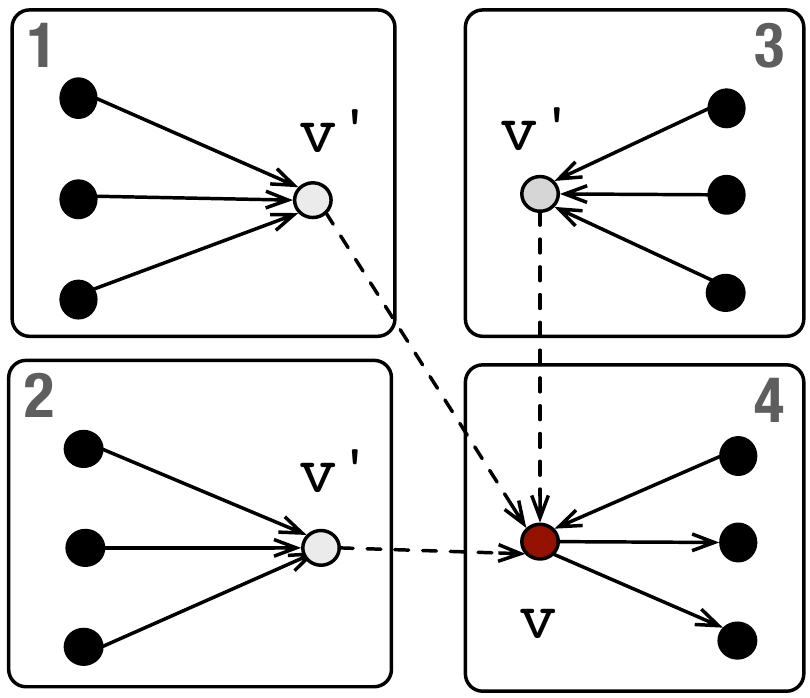}\
		\label{subfig:agent-graph}
	}
	\hspace{0.3cm}
	\subfloat[Vertex-cut]{
		\includegraphics[scale=0.28]{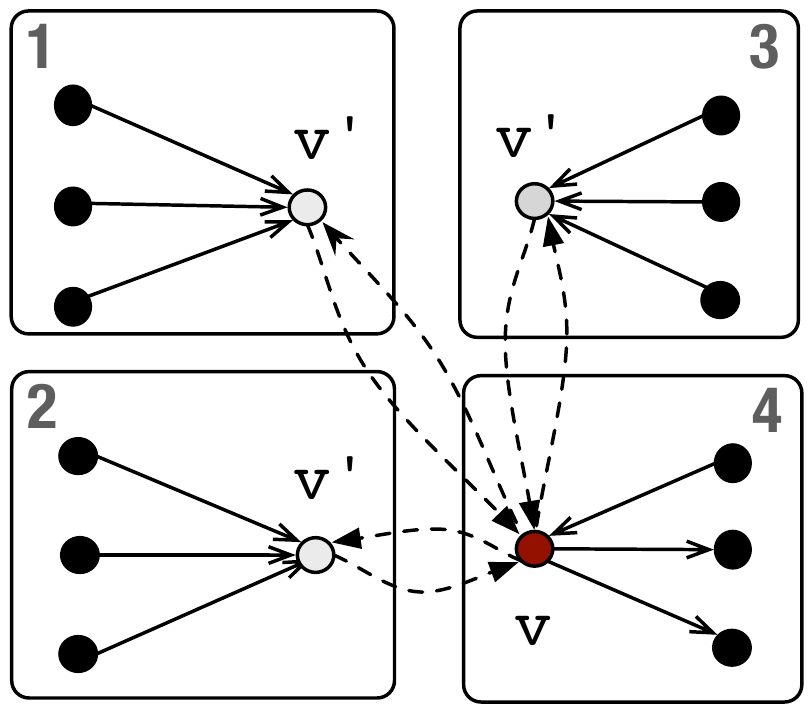}
		\label{subfig:vertex-cut}
	}
\caption{Comparison of Agent-graph and Vertex-cut. The dashed lines represent communication operations.}
\label{fig:compare-graph-models}
\vspace{-0.25cm}
\end{figure}

Note that {\tt vertex-cut} model in PowerGraph is another way to address distributed placement of scale-free graphs. In {\tt vertex-cut}, a vertex can be cut into multiple replicas distributed in different machines, where the remote replica of vertex is called {\em mirror}. Both {\em agent} and {\em mirror} are based on vertex factorization. Conceptually, {\tt agent-graph} can be built from {\tt vertex-cut} partitions by simply splitting one {\em mirror} into one {\em scatter} and one {\em combiner}. However, their mechanisms are fundamentally different. Fig.\ref{fig:compare-graph-models} shows an example to illustrate difference of {\tt agent-graph} and {\tt vertex-cut}. We argue that {\em agent} has obvious advantage over {\em mirror} for expressing message model on directed graphs.

First, {\em agent} has no overhead on maintaining its data consistency with {\em master} while {\em mirror} has to periodically synchronize data with {\em master}. This is because {\em mirror} holds an integrated copy of its {\em master}'s runtime states, while {\em agent}, as comparison, is a message proxy that can only temporally cache and forward messages in single direction. Thus, for traversal-based algorithms on directed graphs, {\tt agent-graph} has much less communication than {\tt vertex-cut}.

Second, the communication cost of {\em agent} is lower than {\em mirror}'s in most of cases. In {\tt vertex-cut}, each {\em master} first accumulates all its {\em mirror}s' data, and then sends the new result to all its {\em mirror}s. Thus, the communication is $2*R$ (R is the total number of all vertices' replicas). In {\tt agent-graph}, one {\em agent} only involves in one direction message delivery, either receive ({\em scatter}) or send ({\em combine}). Thus, it has less communication cost since $|V_s|+|V_c| \leq 2*max (|V_s|, |V_c|) = 2*R$. Take the example in Fig.~\ref{fig:compare-graph-models}. In {\tt vertex-cut} the {\em master} vertex $v$ has to collect all changes of its {\em mirror}s and then spread the newest value to them, while in {\tt agent-graph} $v$ only receives messages from its {\tt agent}s ({\tt combiner}s here) and has no need to update {\tt agent}s. As shown by the number of dashed lines, {\tt agent-graph} requires just half communication of that in {\tt vertex-cut}.

\subsection{Partition Algorithms}
\label{sub:graphPar}
On {\tt Agent-Graph} model, we thoroughly investigated various streaming and 2-pass semi-streaming partitioning methods in~\cite{agentPar}. With the agent-extension, both traditional edge-cut and vertex-cut partitioning methods perform much better for scale-free graphs. In this paper, we only give the pure vertex-cut approach and a streaming partitioning method adapted from PowerGraph's greedy vertex-cut heuristics.

In a pure vertex-cut model, none of edges in original graph $G$ is cut. All cut edges are $G^{\alpha}$ extended edges, i.e. \{$master\rightarrow scatter$\} or \{$combiner \rightarrow master$\} . The extended edges represent communication overhead while original edges represent computational load. We construct {\tt Agent-Graph} by loading edges from the original graph. Assuming that the goal is $k$ partitions, we formalize the objective of k-way balanced partition objective as follow:
\begin{equation}
\begin{split}
&\operatorname*{min}_{A} \frac{1}{k}\sum_{v \in V}{(|S(v)| + |C(v)|)}\\
s.t. \quad &\operatorname*{max}_{1 \leq i < k} | \{e \in E | P(e) = i\} | < (1+\epsilon) \frac{|E|}{k}
\end{split}
\end{equation}
where $\epsilon$ is an imbalance factor, $S(v)$ and $C(v)$ are sets of {\em scatter}s and {\em combiner}s of vertex $v$ respectively.

The loader reads edge list in a stream way and greedily places an edge $(u, v)$ to the partition which minimizes number of new added {\em agents} and keeps edge load balance. The current best $idx$ of partition is calculated by the following heuristic:
\begin{equation}
idx = \operatorname*{arg\,max}_{1 \leq i < k} \{f (u,i) + g(v, i) + \frac{Max - Ne(i)}{\Delta + Max - Min}\}
\end{equation}
, $where$ $\Delta = 1.0$ and
\begin{equation}
f(u, i) = \begin{cases}
	1 & \text{partition $i$ has edges with source $u$}\\
	0  & others
	\end{cases}
\end{equation}
\begin{equation}
g(v, i) = \begin{cases}
	1 & \text{partition $i$ has edges with target $v$}\\
	0  & others
	\end{cases}
\end{equation}
\begin{equation}
Ne(i) = | \{e \in E | P(e) = i\}|
\end{equation}
\begin{equation}
Max = \operatorname*{max}_{1 \leq i < k} Ne(i),\quad
Min = \operatorname*{min}_{1 \leq i < k} Ne(i).
\end{equation}

In default, each machines independently loads a subset of edges, partitions them into $k$ parts, and finally sends remote partitions to their owner machines. During the procedure, machines don't exchange information of heuristic computing. This is the same with the {\em oblivious} mode in PowerGraph. Also, {\tt GRE} supports the {\em coordinated} mode of PowerGraph, where partitioning information are periodically synchronized among all machines.
	
\section{Runtime System Design}
\label{sec:runtime}
{\tt GRE}'s abstractions, {\tt Scatter-Combine} computation model and distributed {\tt Agent-Graph} model, are built on the runtime layer. The runtime system is designed for contemporary distributed systems in which each single machine has multiple multi-core processors sharing memory and is connected to other machines with high performance network. {\tt GRE} follows an owner-compute rule, that launches one single process on each machine and assigns a graph partition to it. Within each machine, the process has multiple threads: one master thread in charge of inter-process communication and multiple worker threads that are forked and scheduled by the master thread to do actual computation. Now, we describe the runtime design from three aspects, with an emphasis on how it bridges {\tt GRE} abstractions and the underlying platform.

\subsection{Data Storage}
\label{sub:data_storage}
From the view of users, graph-parallel applications run on a directed property graph where each vertex is identified by an unique 64-bit integer. Internally, however, {\tt GRE} stores runtime graph data in a distributed way. {\tt GRE} manages three types of in-memory data: graph topology, graph property and runtime states.

\subsubsection{Graph Topology}
Each machine storers a partition of {\tt agent-graph}. The local topology storage is compact and highly optimized for fast retrieve. It includes three parts: graph structure, vertex-id index and agent-extended edges. First, the graph structure stores all assigned ordinary edges in the CSR (Compressed Sparse Row) format~\cite{csr}, where all vertices are renumbered with local 32-bit integer IDs. Local vertex IDs are assigned by the following rule. Assuming that there are $n$ local vertices (i.e. {\em master}), {\em master}s are numbered from 0 to $n$-$1$ in order. Both {\em scatter}s and {\em combiner}s are then continuously numbered from $n$. Second, the vertex-id index provides bidirectional translation between local ID and global ID for all vertices. Third, {\tt GRE} stores agent-extended edges implicitly. For any $master$$\rightarrow$$scatter$ type edge, the $master$ induces it by retrieving a data structure recording all machines that hold its {\em scatter}s. For any $combiner$$\rightarrow$$master$ type edge, the {\em combiner} induces it by local-to-global vertex-id index.
\REM{
\begin{figure}[h]
	\centering	
	\includegraphics[scale=0.4]{number_rule}
\caption{Local vertex ID numbering rule.}
\label{fig:number_rule}
\vspace{-0.25cm}
\end{figure}
}

\subsubsection{Graph Property}
Graph property (i.e. meta data associated with vertices and edges) is decoupled with graph topology. It is separately stored in the Column-Oriented Storage ({\tt COS})~\cite{cos} approach. In {\tt COS}, each type of graph property is stored as a flat array of data items. The local vertex (edge) ID serves as primary key and can directly index the array. For example, in a social network, given one person's {\em local} ID, say $i$, to retrieve his/her {\em name} we directly locate the information by {\em name}[$i$]. {\tt COS} provides fast data load/store between disk and memory, as well as optimizations like streaming data compression. With {\tt COS}, {\tt GRE} can load or store arbitrary types of graph property in need, and run multiple ad-hoc graph analysis continuously.

\subsubsection{Runtime States}
Vertex runtime states play a crucial role in implementing {\tt GRE}'s {\tt Scatter-Combine} computation model. Like graph property, runtime states are stored in flat arrays and indexed by local vertex ID. Conceptually, there are three types of runtime vertex states:
\begin{itemize*}
\item {\em vertex\_data} is the computing results only owned by {\em master} vertex, and updated by {\em apply}.
\item {\em scatter\_data} is the data that one vertex wants to {\em scatter} by messages, owned by {\em master} and {\em scatter}.
\item {\em combine\_data} is the data on which an active message executes {\em combine}, owned by {\em master} and {\em combiner}.
\end{itemize*}

\vspace{-0.25cm}
\begin{figure}[h]
	\centering	
	\subfloat[Runtime States Vector]{
		\label{subfig:data-types}
		\includegraphics[scale=0.35]{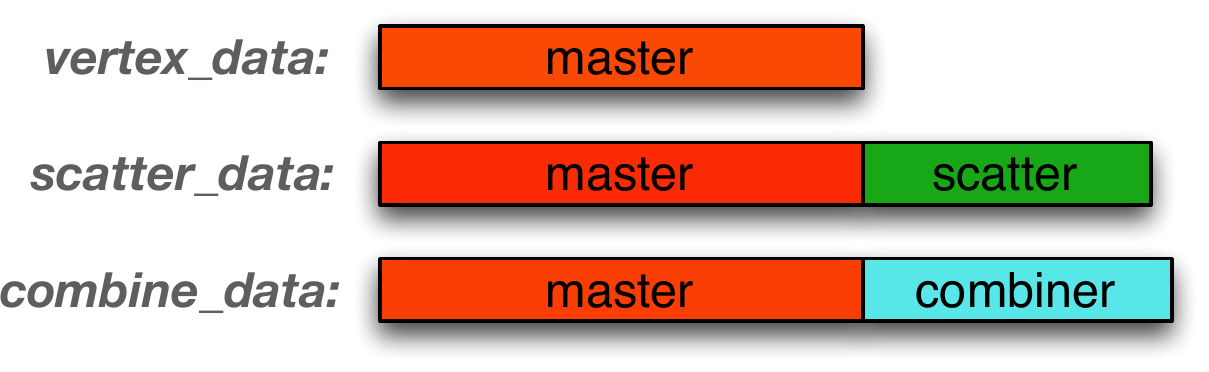}
	}
	\\
	\subfloat[Runtime States in Agent-Graph]{	
		\label{subfig:states-agent}
		\includegraphics[scale=0.35]{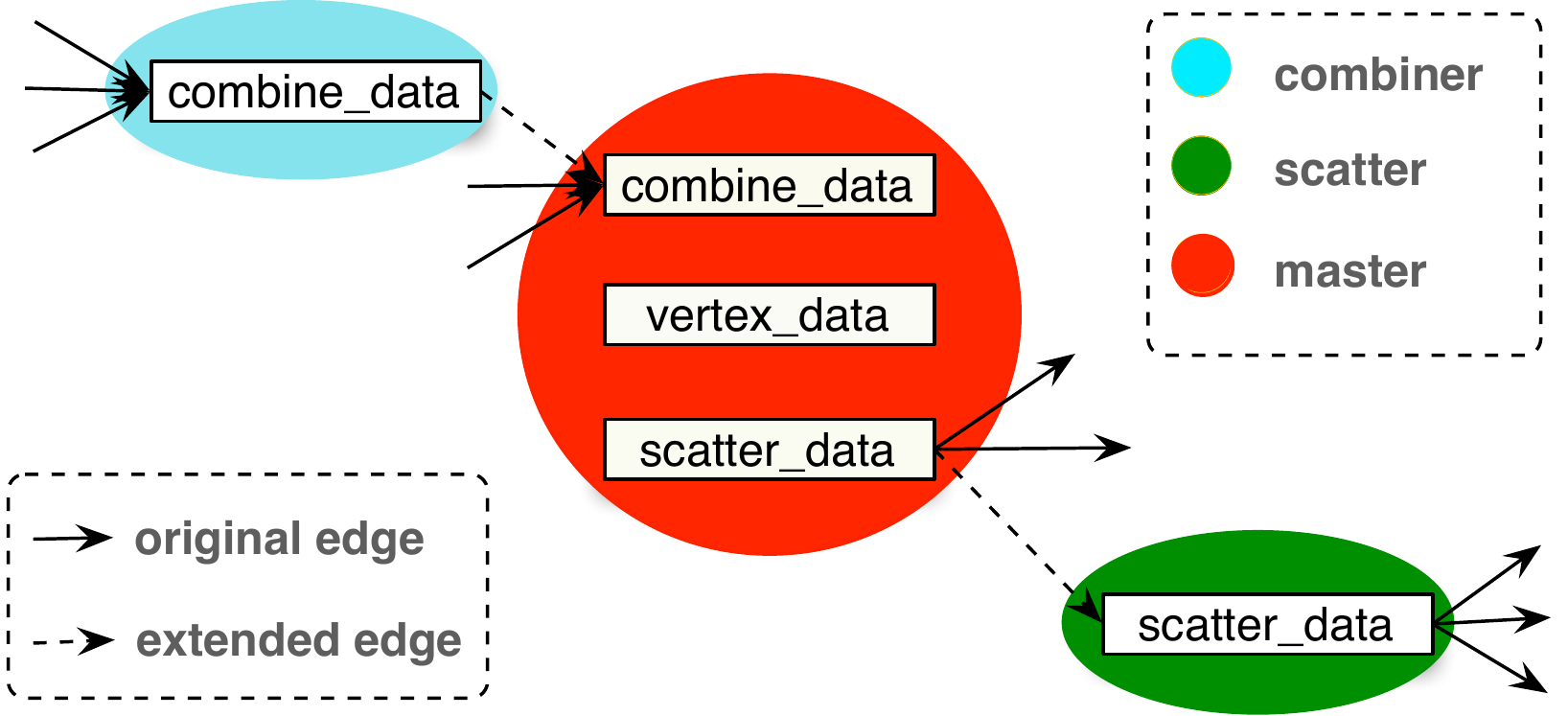}
	}
\caption{Illustration of Runtime States.}
\label{fig:compute_illustration}
\vspace{-0.25cm}
\end{figure}

Tabel.~\ref{tab:runtime-states} is the runtime state setting of algorithms in Fig.~\ref{fig:applications}. Note that for performance optimization, {\tt GRE} allows the {\em vertex\_data} vector refer to {\em scatter\_data} or {\em combine\_data}. However, for data consistency, {\tt GRE} requires {\em scatter\_data} and {\em combine\_data} be physically different. 
\vspace{-0.20cm}
\begin{table}[h]
\caption{Runtime States Setting in Examplary Algorithms}
\centering
{\small
\begin{tabular}{|c|l|l|l|}
\hline
Algorithm & {\em scatter\_data} & {\em scatter\_data} & {\em combine\_data}\\
\hline
PageRank & {\tt pr} & {\tt pr} & {\tt pr\_combine}\\
SSSP & {\tt distance} & {\tt oldDistance} & {\tt distance}\\
CC & {\tt label} & {\tt oldLabel} & {\tt label}\\
\hline
\end{tabular}
}
\label{tab:runtime-states}
\vspace{-0.25cm}
\end{table}

Data consistency of vertex states is automatically maintained by the specification of {\tt Scatter-Combine} primitives. (1) For an ordinary vertex({\em master}), its {\em scatter\_data} can only be updated by initialization or {\em apply}, while for a {\em scatter-agent} its {\em scatter\_data} can only be updated by the message from its {\em master}. During {\tt scatter-combine} phase, the {\em scatter\_data} doesn't change, and is valid only when the vertex is active for {\em scatter}. (2) For a vertex, either {\em master} or {\em combiner-agent}, {\em combine} operation may change and can only change its {\em combine\_data}. If the vertex is a $master$, {\em combine} on it incurs a future {\em apply}. If the vertex is a {\em combiner-agent}, in the future it will send an active message to its remote {\em master} and then reset its {\em combine\_data}. (3) During {\tt apply} phase, each active {\em master} executes an {\em apply} in which it updates the {\em vertex\_data}, optionally recomputes {\em scatter\_data}, and resets the {\em combine\_data}. 

\subsection{Active Messages}
Active message hides underlying details and difference of intra-machine shared-memory and inter-machine distributed memory. With one-sided communication and fine-grained data synchronization, {\tt GRE}'s runtime layer provides efficient support to active messages.

\subsubsection{One-sided Communication}
A daemon thread (master of local process) keeps monitoring the network and receiving incoming data, meanwhile sending data prepared by {\em worker}s. {\tt GRE} supports asynchronous communication of multiple message formats simultaneously.

\begin{figure}[h]
	\centering	
	\includegraphics[scale=0.35]{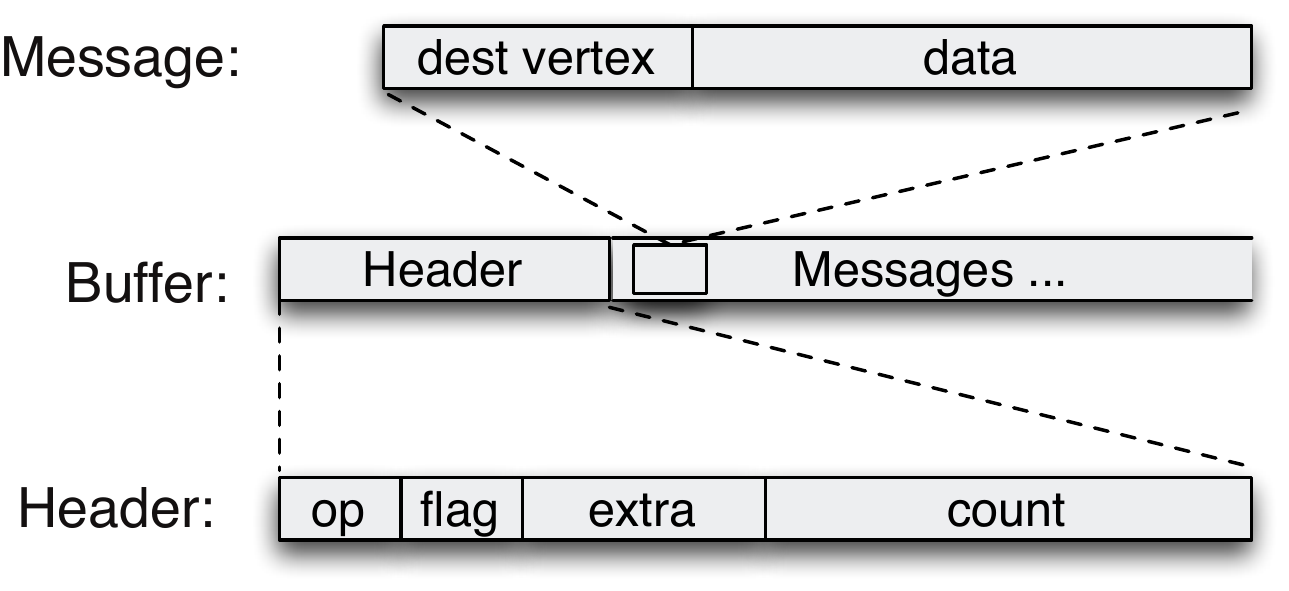}
\caption{Format of Communication Buffer.}
\label{fig:comm_protocol}
\end{figure}

The communication unit is a memory block whose format is shown in Fig.\ref{fig:comm_protocol}. It consists of two parts, header and messages. The header is a 64-bit structure that implements a protocol to support user-defined communication patterns. Messages with the same format are combined into one buffer and the format registration information is encoded in the header. The fields {\em op}(8 bits) and {\em flag}(8 bits) decide what actions to take when receive the message. The filed {\em count}(32 bits) is the number of messages in the buffer. Message is vertex-grain, containing destiny vertex and message data. Besides, a buffer with zero message is legal, where the header-only buffer is used to negotiate among processes.

\vspace{-0.3cm}

Each process maintains a buffer pool where the buffer size is predefined globally. One buffer block can be filled with arbitrary messages not exceeding the capacity. As the block has encoded all information in its header, all interactions either between local {\em master} and {\em worker}s or among distributed processes are one-sided and asynchronous.

\subsubsection{Fine-grained Parallelism and Synchronization}
In {\tt GRE}'s master-workers multi-threading mode, all computation ({\em scatter} and {\em combine}) are carried by worker threads in parallel. Note that multiple active messages may do {\em combine} operation on the same vertex simultaneously, which leads to data race. According to our previous statistics~\cite{vlock}, given the sparse and irregular connection nature of real graphs the probability of real conflicts on any vertex is very low. In default, {\tt GRE} uses the proven high performance virtual lock mechanism ({\it vLock}~\cite{vlock}) for vertex-grained synchronization.

{\tt GRE} provides two methods of worker thread organization, i.e. thread pool and thread groups. In thread pool mode, the process maintains a traditional thread pool and a set of virtual locks. All {\em combine} operations are implicitly synchronized by the {\it vLock}. Thread groups is an alternative method,  addressing the issue that for multi-socket machine, frequent atomic operations and data consistency across sockets may lead to high overhead. In thread groups mode, worker threads are further divided into groups. Each thread group runs on one socket, computes on one of local vertices' disjoint subsets, and communicates with other groups by FastForward channels~\cite{fastforward}. Besides, each thread group has an independent set of {\it vLock} privately.

\subsection{Fault Tolerance}
Fault tolerance in {\tt GRE} is achieved by checkpointing, i.e. saving snapshot of runtime states periodically in a given interval of super steps. The process is similar to that in Pregel but much simpler. During checkpointing, {\tt GRE} only needs to backup for native vertex runtime states and active vertex bitmap, abandoning all {\tt agent} data and temporal messages. Besides, thanks to the column-oriented-storage, {\tt GRE} can dump and recover runtime data image very fast. For {\tt GRE}, since it keeps all runtime data in-memory, failures rarely happen and typically incurred by message loss over network which can be caught by a communication component at the end of a super-step.

\section{Experimental Evaluation}
\label{sec:evaluation}

\begin{figure*}[ht]
	\centering
	\subfloat[PageRank on Twitter]{
		\includegraphics[scale=0.29]{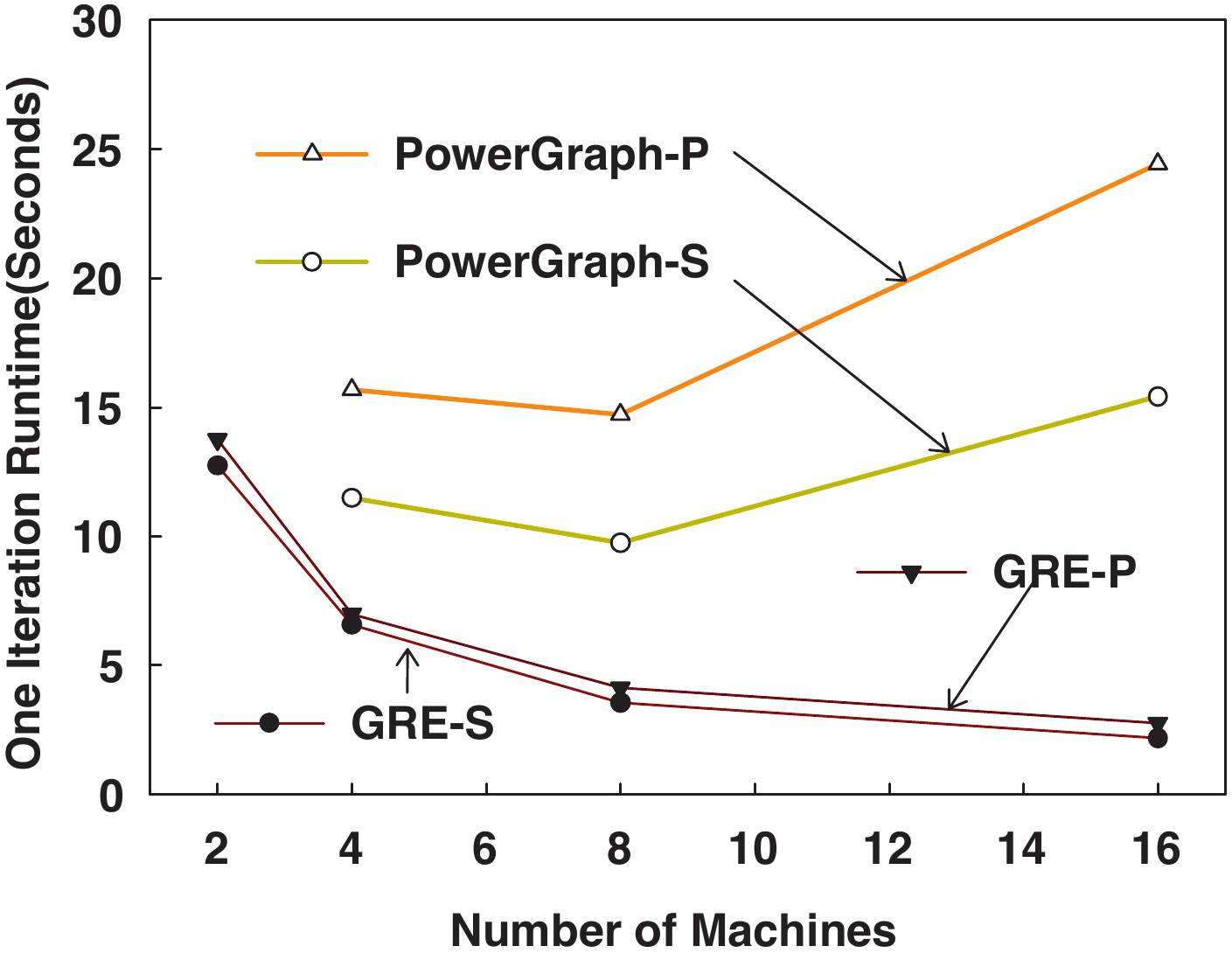}
		\label{subfig:twitter-pr-runtime}
	}
	\subfloat[PageRank on SK-2005]{
		\includegraphics[scale=0.29]{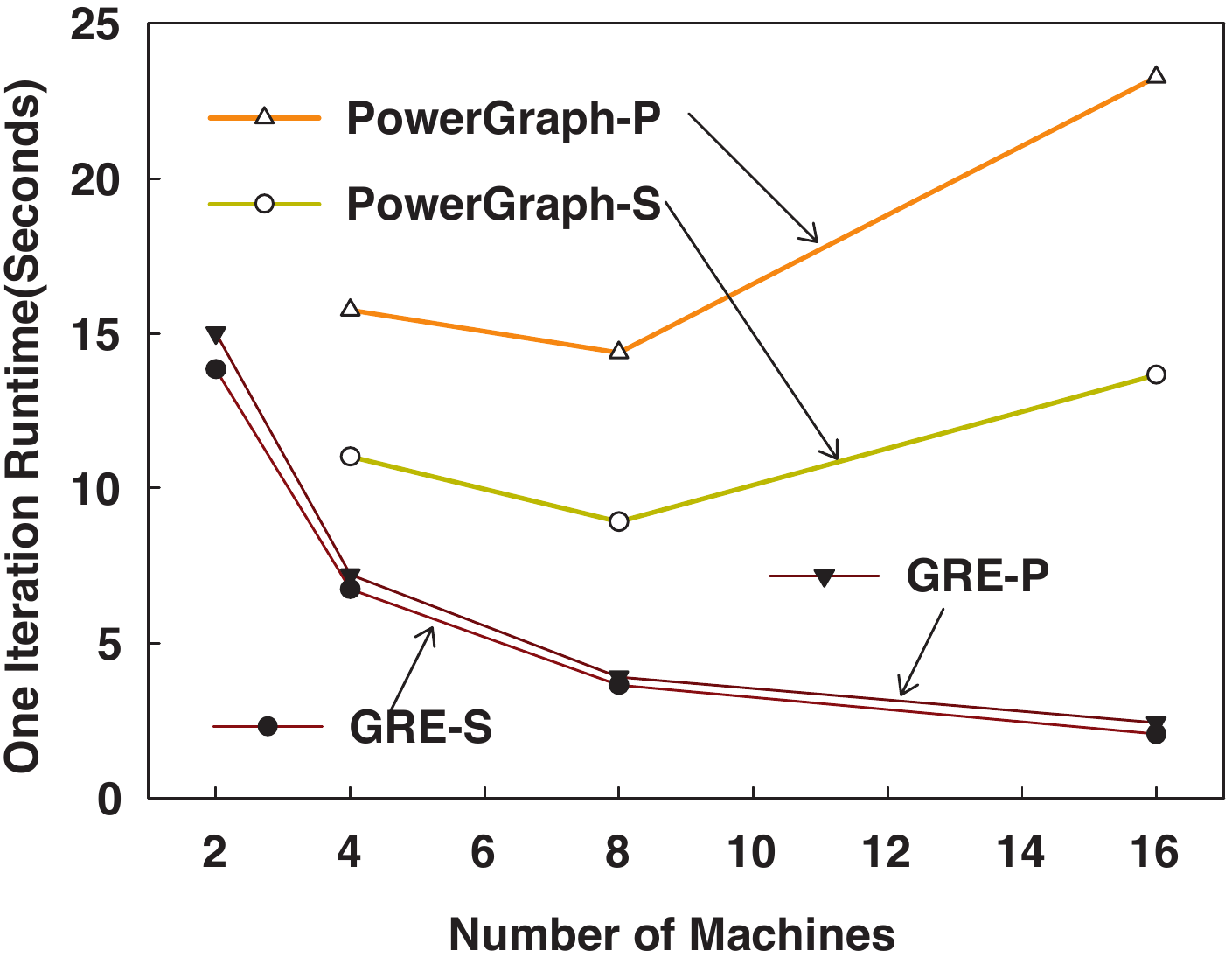}
		\label{subfig:sk-pr-runtime}
	}
	\subfloat[SSSP on Twitter]{
		\includegraphics[scale=0.29]{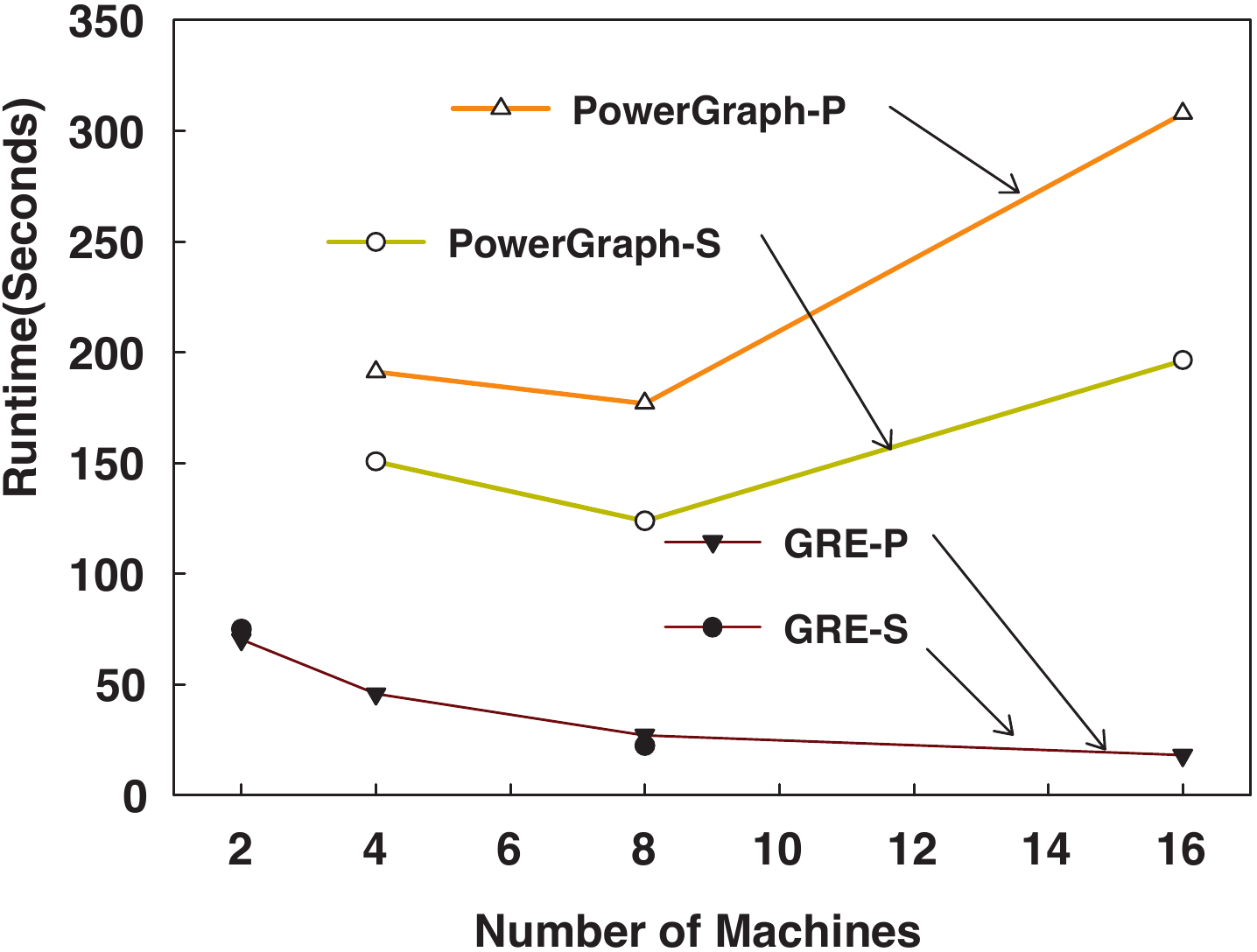}
		\label{subfig:twitter-sssp-runtime}
	}
	\subfloat[CC on Graph500-27]{
		\includegraphics[scale=0.29]{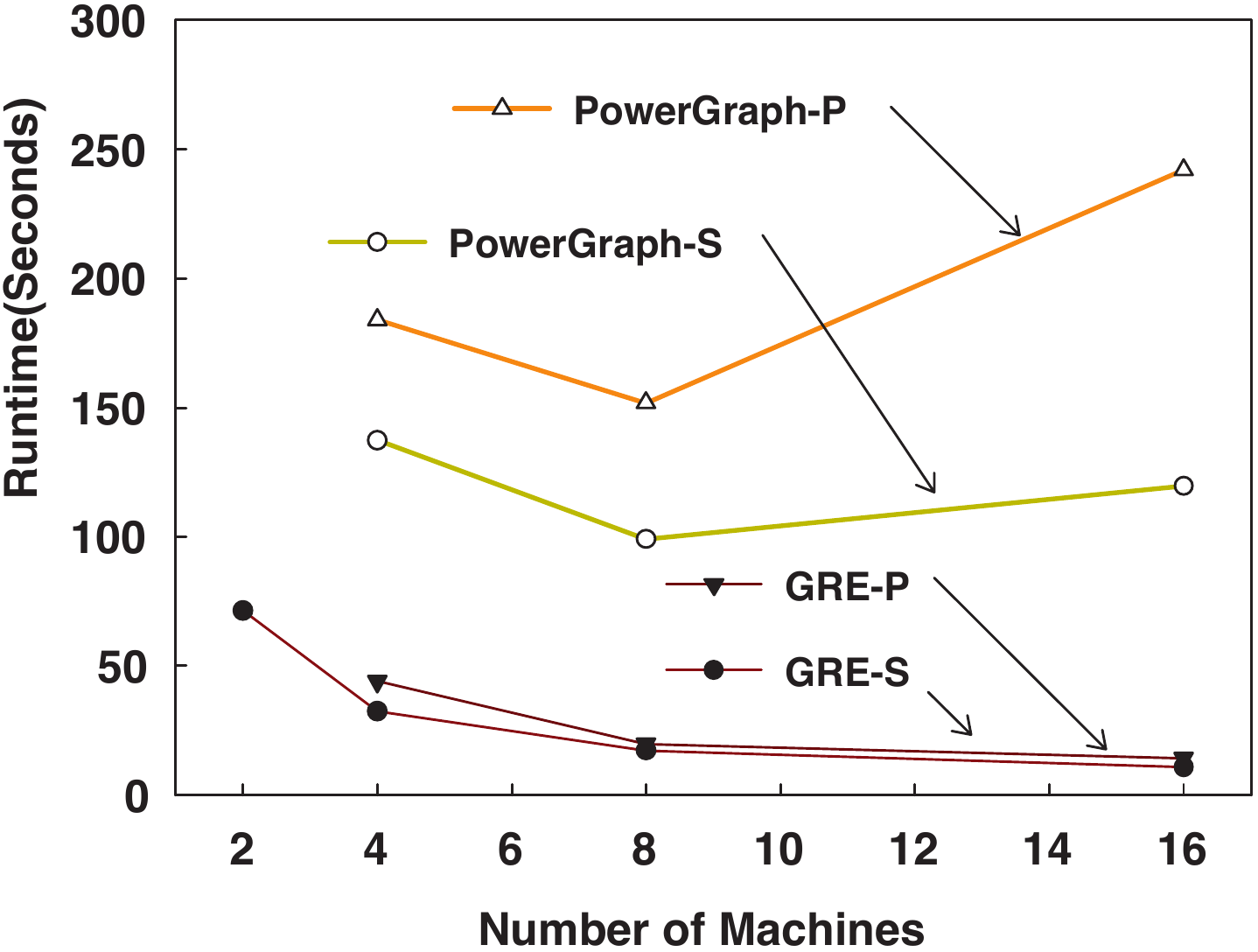}
		\label{subfig:27-08-cc-runtime}
	}
\caption{Run-time of GRE and PowerGraph. The Graph500-27 is generated by Graph500, with 128M vertices and 2B edges. }
\label{fig:runtime}
\vspace{-0.5cm}
\end{figure*}

\begin{figure*}[ht]
	\centering
	\subfloat[PageRank]{
		\includegraphics[scale=0.28]{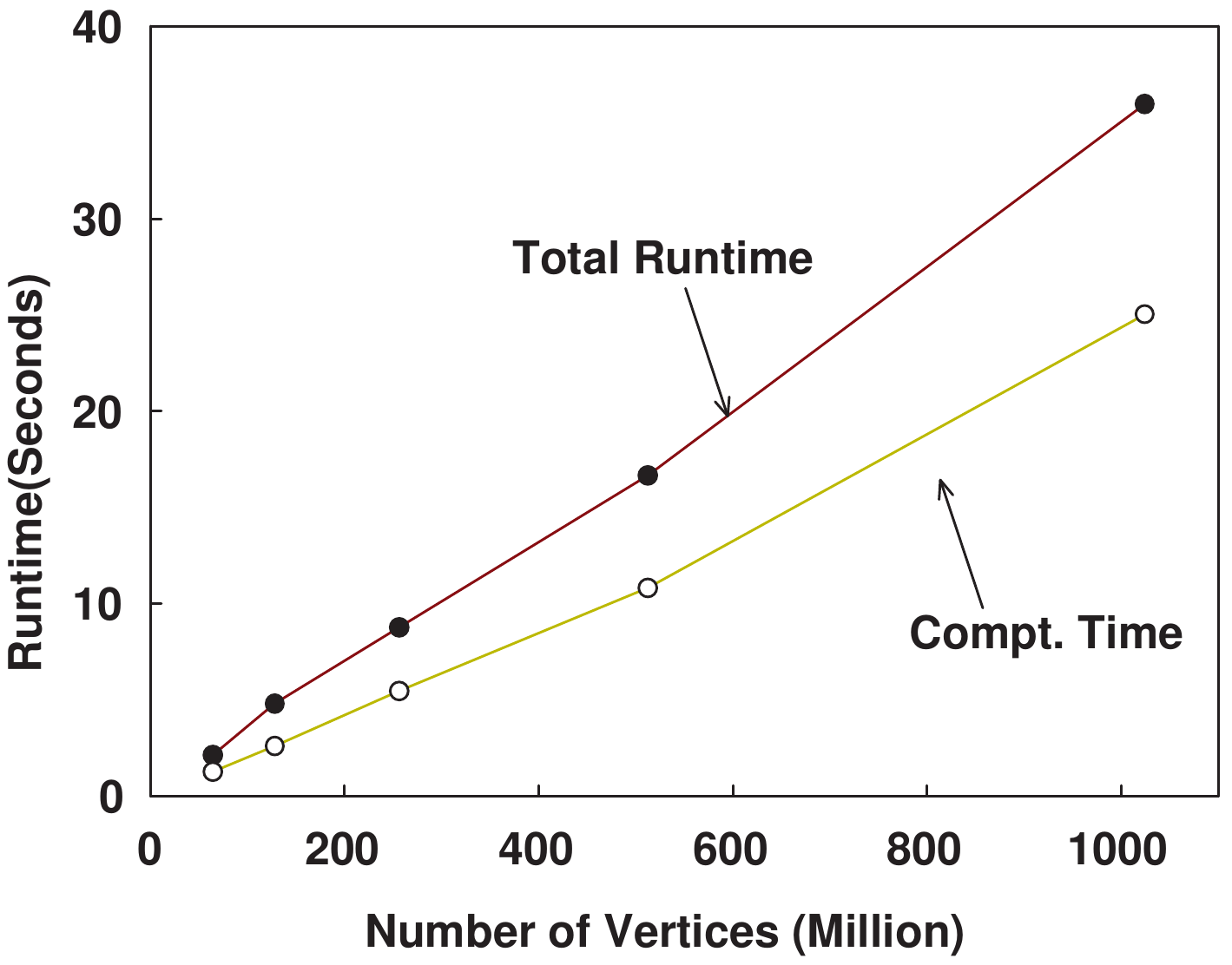}
		\label{subfig:graphs-pr}
	}
	\subfloat[SSSP]{
		\includegraphics[scale=0.28]{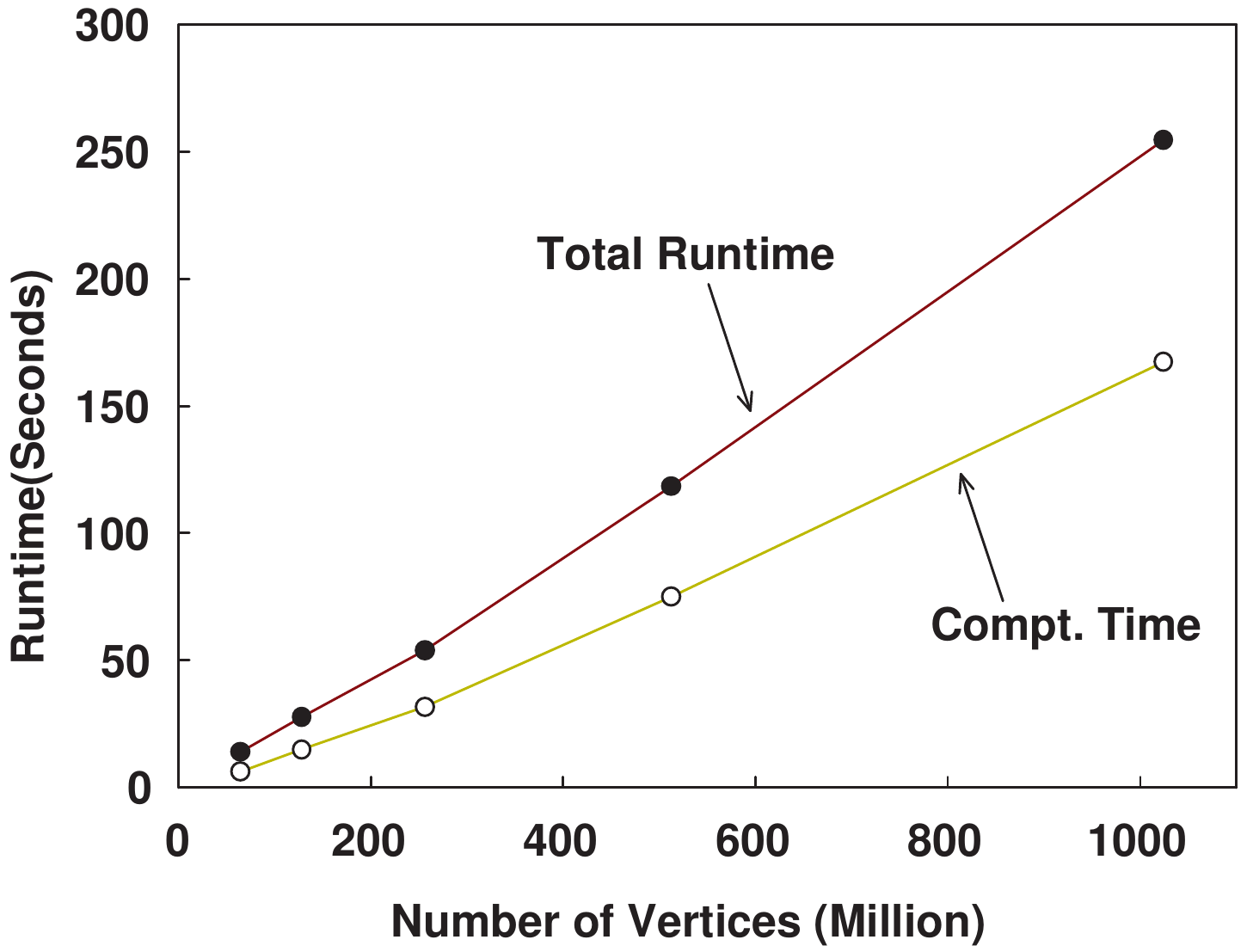}
		\label{subfig:graphs-sssp}
	}
	\hspace{0.25cm}
	\subfloat[Connected Components]{
		\includegraphics[scale=0.28]{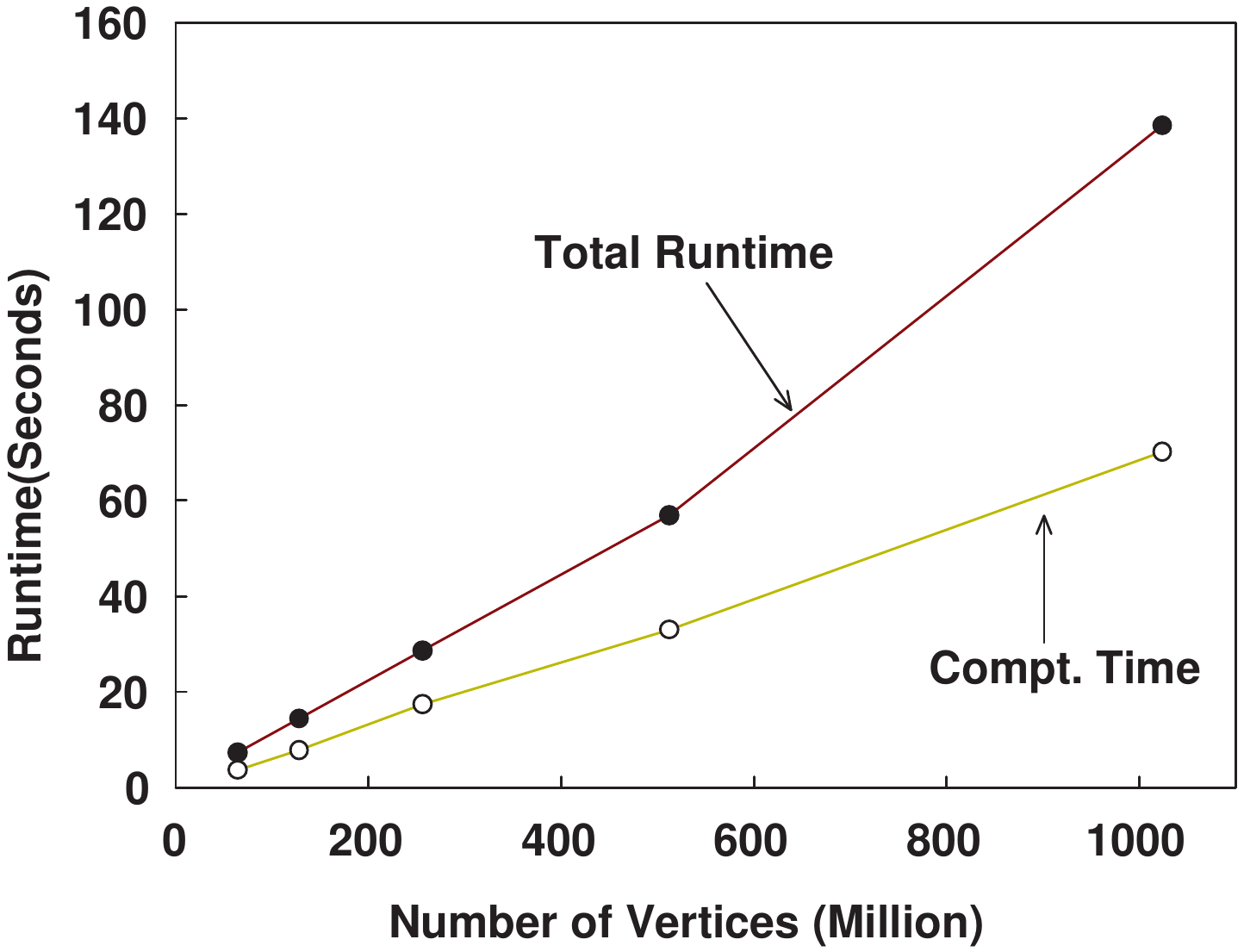}
		\label{subfig:graphs-cc}
	}
\caption{Scalability of GRE(-P) over graph sizes. The graphs are generated by Graph500, with mean out-degree 16 and varying vertex numbers from 64M to 1B. Edge weights are randomly generated from [1, 65535]. We run {\tt GRE-P} on 16 machines.}
\label{fig:scalability-problem-sizes}
\vspace{-0.25cm}
\end{figure*}

\begin{descdesp}
\item[{\em Platform.}]  The experimental platform is a 16-node cluster. Each node has two six-core Intel Xeon X5670 processors, coupled with 48GB DDR-1333 RAM. All nodes are connected with MLX4 Infiniband network of 40GB/s bandwidth. The operating system is Suse server Linux 10.0. All applications of both {\tt GRE} and PowerGraph (GraphLab 2.2) are compiled with GCC 4.3.4 and OpenMPI 1.7.2.

\item[{\em Benchmark and Datasets.}] We choose three representative algorithms---PageRank, Single Source Shortest Path(SSSP) and Connected Components(CC). We use 9 real-world and a set of synthetic graph datasets. The real-world datasets we used are summarized in Table.~\ref{datasets}. To the best of our knowledge, this set includes all available largest graphs in public. The synthetic graphs are R-MAT graphs generated using Graph500 benchmark with parameters a=0.57, b=c=0.19 and d=0.05. They have fixed out-degree 16, and varying numbers of vertices from 64 million to 1 billion.
\begin{table}[h]
\caption{Summary of Real Graph Datasets.}
\centering
{\small
\begin{tabular}{|l||c|c|c|}
\hline
Name & $|V|$ & $|E|$ & Type\\
\hline
LiveJournal\cite{lj} & 5,363,260 & 79,023,142 & Social\\
Hollywood\cite{hollywood} & 2,180,759 &	228,985,632 & Social\\
Orkut\cite{orkut} &	3,072,626 & 234,370,166 & Social\\
Arabic\cite{sk} & 22,744,080 & 639,999,458 & Web\\
Twitter\cite{twitter} & 41,652,230 & 1,468,365,182 & Social\\
Friendster\cite{friendster} &65,608,366 &1,806,067,135 &Social\\
SK\cite{sk} & 50,636,154 & 1,949,412,601 & Web \\
UK\cite{uk} & 105,896,555 & 3,738,733,648 & Web\\
AltaVista\cite{altavista} & 1,413,511,390 & 6,636,600,779 & Web\\
\hline
\end{tabular}
}
\label{datasets}
\end{table}
\vspace{-0.25cm}

\item[{\em Notation.}] Since graph partitioning strategies are closely related to parallel performance, we implemented two graph partition settings on {\tt GRE} when comparing to {\tt PowerGraph}. Table.~\ref{tab:partition} gives the name notation of different partition strategies.  The graph partitioning is evaluated in terms of both {\it equivalent edge-cut rate} and {\it cut-factor}. The {\it edge-cut rate} is defined as the rate of communication edges count to total number of edges, while the {\it cut-factor} is the rate of communication edge count over total number of vertices.
\begin{table}[h]
\caption{Framework Configuration}
\centering
{\small
\begin{tabular}{|l|l|}
\hline
Name & Graph Ingress Setting\\
\hline
GRE-S & Serial loading and partitioning\\
GRE-P & Parallel loading and partitioning\\
PowerGraph-S & Serial load and partitioning (Coordinated) \\
PowerGraph-P & Parallel load and partitioning (Oblivious)\\
\hline
\end{tabular}
}
\label{tab:partition}
\end{table}

\item[{\em Result Summary.}] Our evaluation focuses on {\tt GRE}'s performance and scalability on different machine scales and problem sizes. The results of three benchmark programs and graph partitioning are summarized as following.
\begin{desclist}
\item {\tt GRE} achieves good performance on all three benchmark programs, 2.5$\sim$7.6 (6.6$\sim$17.0) times better than {\tt PowerGraph} when running on 8 (16) machines (Fig.~\ref{fig:runtime}). Specifically, compared to other systems, {\tt GRE} achieves the best performance for PageRank on Twitter graph (Table.~\ref{tab:pr-relative-perf}).
\item {\tt GRE} can efficiently scale to either hundreds of CPU cores (Fig.~\ref{fig:runtime}) or billions of vertices (Fig.~\ref{fig:scalability-problem-sizes}). {\tt PowerGraph}, however, can scale to neither 16 machines due to communication overhead, nor the synthetic graph with 512 million vertices and 8 billion edges due to its high memory cost.
\item {\tt GRE} shows significant advantage on graph partitioning. Compared to random hash method, {\tt GRE}'s {\tt agent-graph} can partition 9 real-world graphs with 2$\sim$11 times improvement on equivalent edge-cut (Fig. \ref{subfig:graphs-par-rate}). Compared to {\tt vertex-cut} method in {\tt PowerGraph-S}({\tt P}), when partitioning Twitter and SK-2005 into 4$\sim$16 parts, with the same greedy heuristics {\tt GRE-S}({\tt P}) shows 12\%$\sim$35\% (29\%$\sim$ 58\%) improvement on equivalent edge-cut (Fig.~\ref{subfig:twitter-par}, \ref{subfig:sk-par}).
\end{desclist}
\end{descdesp}

\subsection{Performance Analysis}
\label{sub:bench_perf}
We evaluate {\tt GRE} and compare it with other frameworks in terms of {\em strong scalability} and {\em weak scalability}. In strong scalability test, the problem size (graph size) is constant while the number of machines is increased. In weak scalability test, the problem size increases with a given number of machines.

\begin{figure*}[ht]
	\centering
	\subfloat[PageRank on Twitter]{
		\includegraphics[scale=0.29]{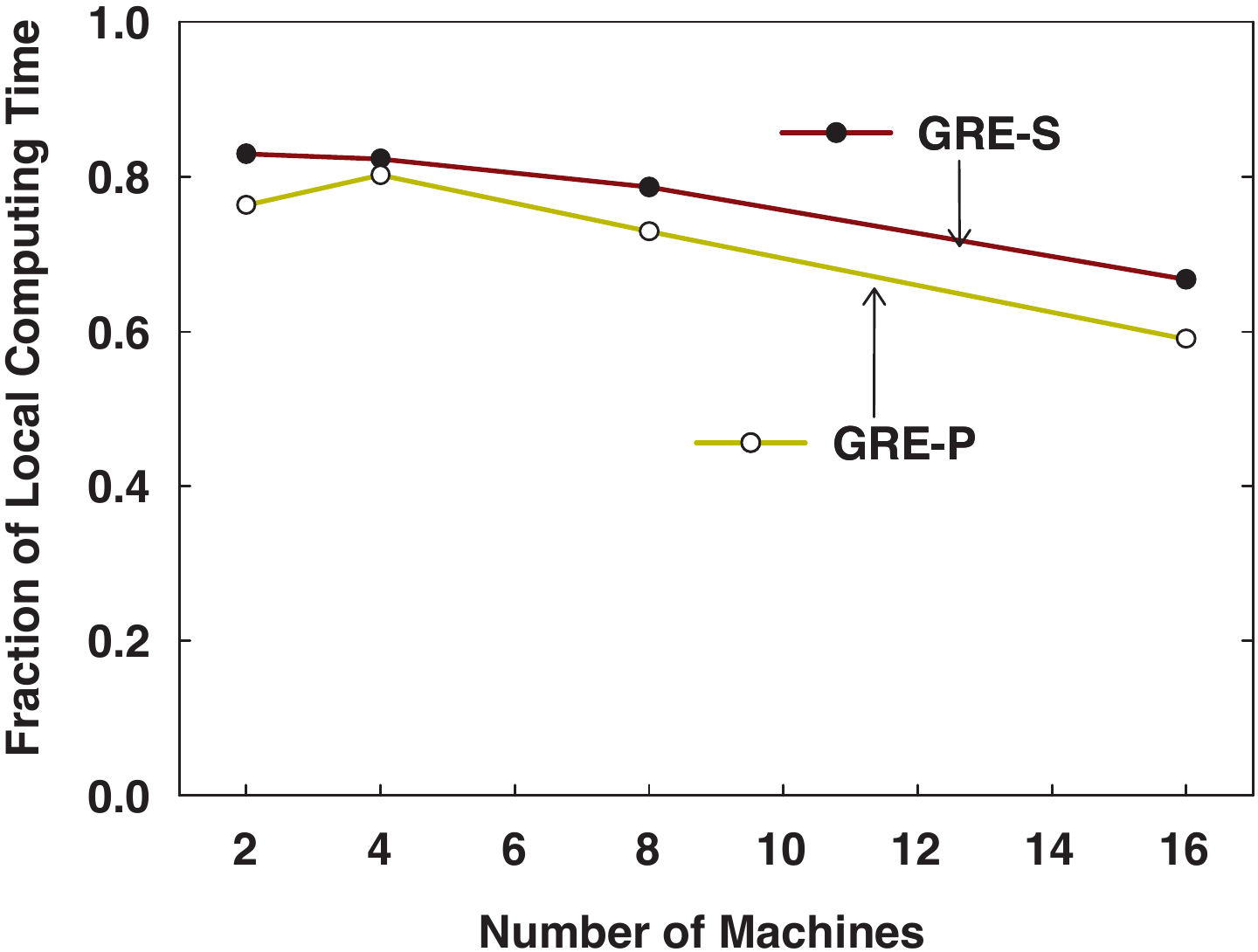}
		\label{subfig:twitter-pr-comp}
	}
	\subfloat[PageRank on SK-2005]{
		\includegraphics[scale=0.29]{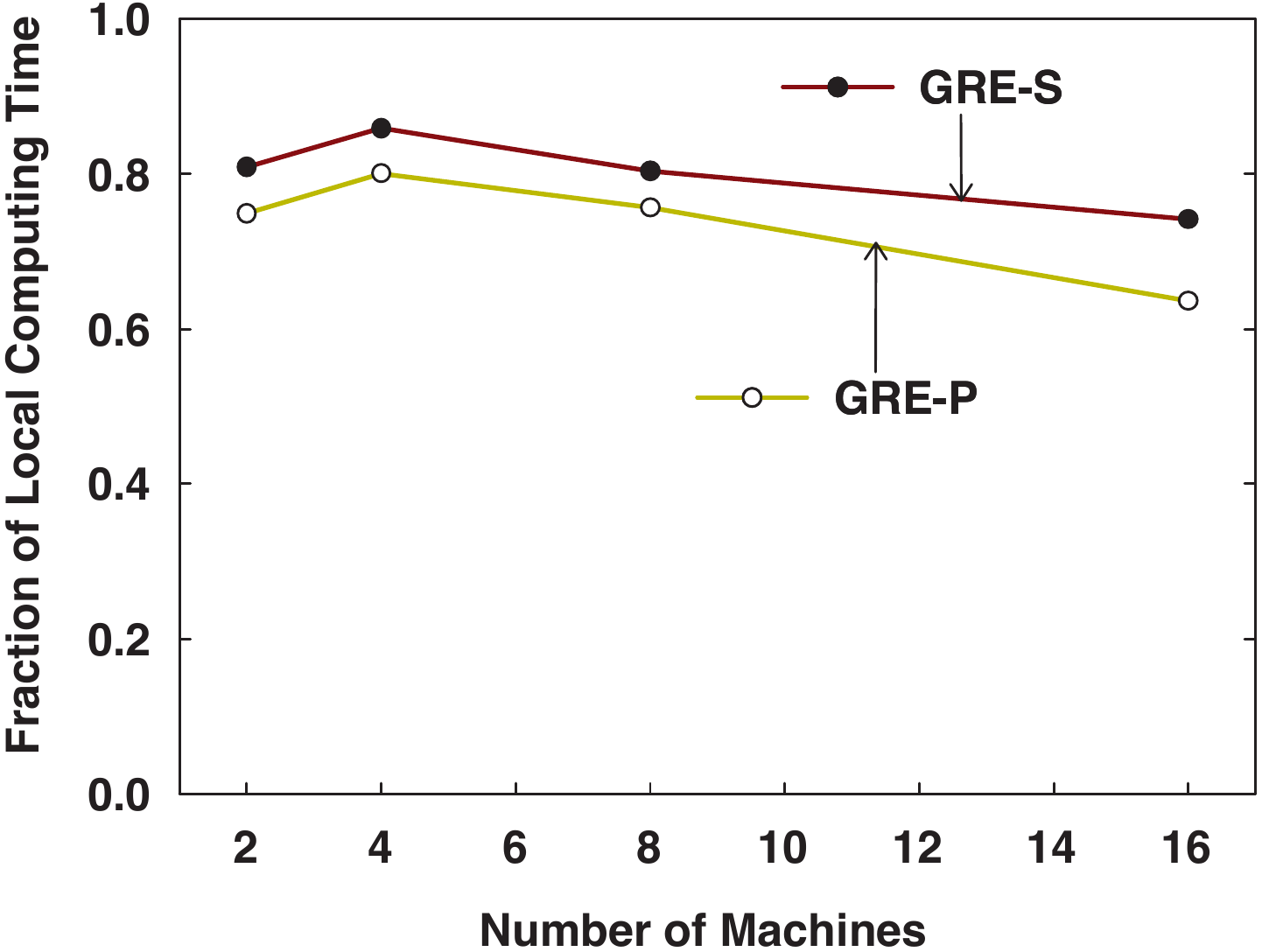}
		\label{subfig:sk-pr-comp}
	}
	\subfloat[SSSP on Twitter]{
		\includegraphics[scale=0.29]{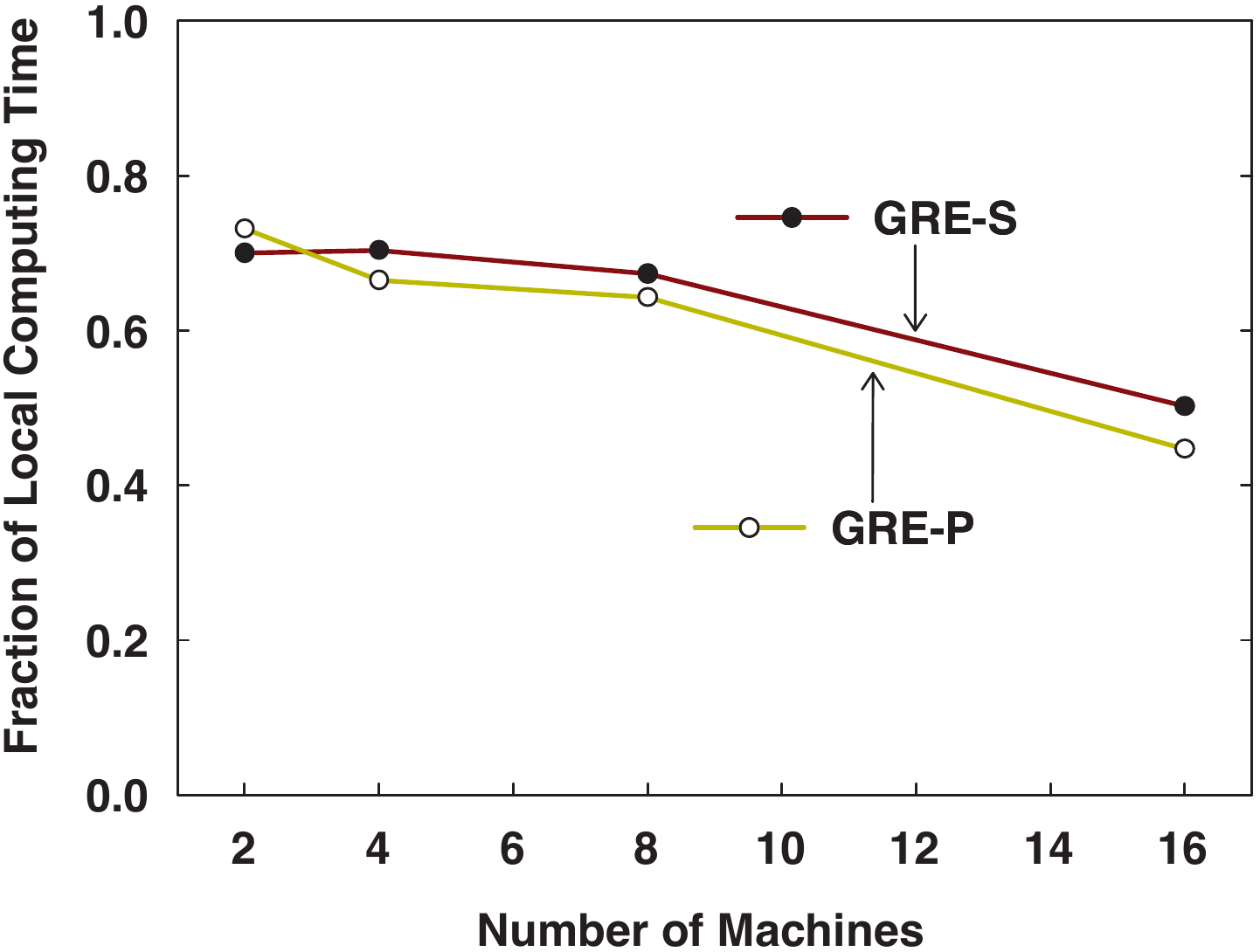}
		\label{subfig:twitter-sssp-comp}
	}
	\subfloat[CC on Graph500-27]{
		\includegraphics[scale=0.29]{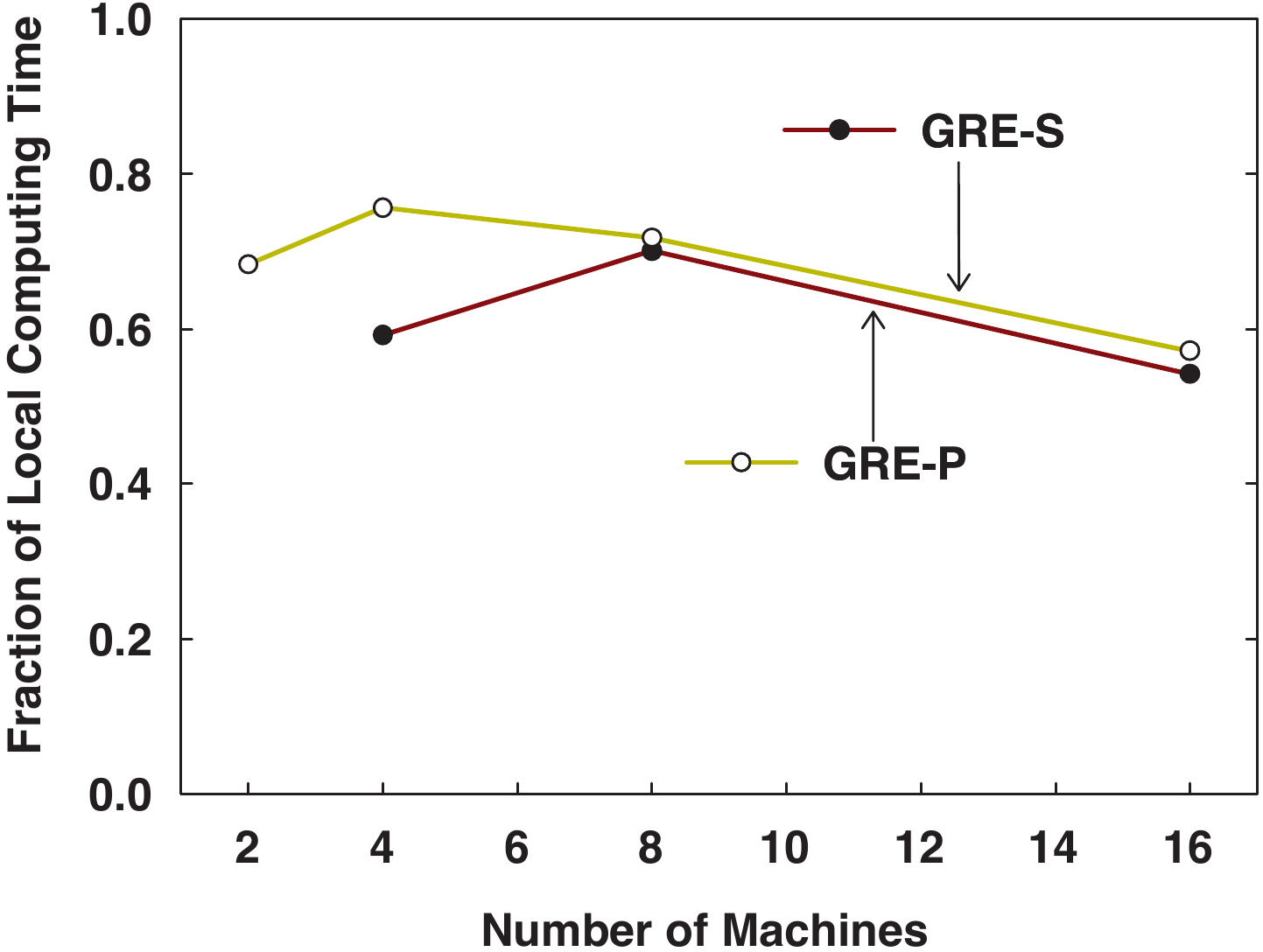}
		\label{subfig:27-08-cc-comp}
	}
\caption{Local Computing Time of GRE. The Graph500-27 is generated by Graph500, with 128M vertices and 2B edges.}
\vspace{-0.25cm}
\end{figure*}

\subsubsection{Strong Scalability}
\label{subsub:perf}
For {\tt PowerGraph}, we adopt the reference implementation in the latest package, with minor modification to ensure its vertex computation same with that in {\tt GRE}. In {\tt PowerGraph}, the PageRank is implemented in a traditional GraphLab way, while SSSP and CC are implemented by emulating {\tt  Pregel}'s {\tt combiner} of messages.
\begin{descdesp}
\item[\em PageRank.] We use two types of real graphs, Twitter social network and SK-2005 web graph, as input graphs. Results of one iteration runtime are shown in Fig.~\ref{subfig:twitter-pr-runtime}  and Fig.~\ref{subfig:sk-pr-runtime} respectively. First, for both graphs, {\tt GRE} overwhelmingly outperforms {\tt PowerGraph} with 1.6$\sim$9.5 times better performance on 2$\sim$16 machines. Second, {\tt GRE} shows nearly linear scalability over increasing machines. In Fig.~\ref{subfig:twitter-pr-runtime}, {\tt GRE-S} and {\tt GRE-P} on 16 machines show the speedup of 5.82 and 5.10 over 2 machines respectively. Similarly in Fig.~\ref{subfig:sk-pr-runtime}, {\tt GRE-S} and {\tt GRE-P} on 16 machines show a speedup of 6.68 and 6.15 over 2 machines.

\item[\em SSSP.] We evaluate it on the Twitter graph whose edge weights are generated by randomly sampling integers from $[1, 65535]$. The program records both distance and predecessor in shortest path for each vertex. Results are shown in Fig.\ref{subfig:twitter-sssp-runtime}. First, {\tt GRE} shows very good performance, e.g. on 16 machines {\tt GRE-S} and {\tt GRE-P} take only 15.46 and 18.08 seconds respectively. Also, like in {\tt PageRank}, {\tt GRE} outperforms {\tt PowerGraph} with significant advantage -- 3.7$\sim$17.0 times better performance on 2$\sim$16 machines. We scrutinize {\tt PowerGraph}'s implementation and find that for traversal-based algorithms (e.g. SSSP and CC) on the directed graph, its graph data model requires much more communication(during the {\tt apply} phase) than that of {\tt GRE}, which leads to the performance loss. Second, {\tt GRE} shows medium speedup. Specifically, {\tt GRE-S} and {\tt GRE-P} on 16 machines show the speedup of 4.85 and 3.89 over 2 machines respectively, lower than that in PageRank.

\item[\em CC.]  We run CC on the synthetic undirected graph (namely, Graph500-27) with 128M vertices and 2B edges, generated by Graph500 benchmark. The results, as shown in Fig.~\ref{subfig:27-08-cc-runtime}, are similar to that of PageRank and SSSP. Specifically, {\tt GRE-P} achieves a speedup of 6.51 on 16 machines over 2 machines, comparable to PageRank and better than SSSP.
\end{descdesp}

{\tt GRE}'s performance benefits from its efficient computation and communication model, that exploits fine-grained parallelism to overlap computation with communication and adopts active message to reduce communication overhead. Compared to PowerGraph's RPC-based multiple-handshake methods, {\tt GRE}' one-sided communication is faster. As we known, parallel scalability is limited by network communication, or in other words, high ratio of computation to communication implies high scalability. We investigate the local computing time of three benchmark programs in {\tt GRE}. As expected, results in Fig.~\ref{subfig:twitter-pr-comp}$\sim$Fig.~\ref{subfig:27-08-cc-comp} show that the ratio of computation time is as high as $45\%\sim80\%$ for the three benchmark programs. That means they can continue to scale well with larger number of machines.

Especially, PageRank on Twitter graph is widely used for performance comparison between different systems in public literatures. We collect these data in Table~\ref{tab:pr-relative-perf}, including {\tt GRE}'s. Compared to other distributed frameworks, {\tt GRE} shows competitive performance.
\vspace{-0.20cm}
\begin{table}[h]
\caption{One iteration run-time of PageRank on Twitter.}
\centering
{\small
\begin{tabular}{|l|l|l|}
\hline
Framework & Time (seconds) & \#Nodes (\#Cores)\\
\hline
GRE-S & 2.19 & 16 (192)\\
GRE-P & 2.77 & 16 (192)\\
PowerGraph\cite{powergraph} & 3.6 & 64 (512)\\
Twister\cite{twister}& 36 & 64 (256) \\
Spark\cite{spark} & 97.4 & 50 (100) \\
\hline
\end{tabular}
}
\label{tab:pr-relative-perf}
\end{table}
\vspace{-0.20cm}
\subsubsection{Weak Scalability}
We further investigate {\tt GRE}'s scalability over problem sizes on given number of machines. Without loss of generality, we consider {\tt GRE-P}. Here we use synthetic graphs generated by Graph500 benchmark~\cite{graph500}. The graph scales from 64M to 1B vertices with the fixed out-degree 16 (may reach 17 in practice). The edge weights are integers sampled from $[1, 65535]$. Due to the limit of memory capacity, the SSSP program only records the distance of each vertex to source.

Run-time of three benchmark programs are shown in Fig. \ref{subfig:graphs-pr}$\sim$\ref{subfig:graphs-cc} . We can see that for all three programs, {\tt GRE} shows excellent scalability on problem sizes, with close to or lower than linear increasing runtime. Specifically, for the largest graph with 1 billion vertices and 17 billion edges, {\tt GRE} can compute one PageRank iteration in 40s, SSSP in 255s, and CC in 139s.

An important phenomena we observed is that {\tt PowerGraph} can not scale to such a large graph size on the 16 machines because the memory consumption exceeds the physical memory capacity (768GB). Compared to {\tt GRE}, {\tt PowerGraph} requires at least 2 times more memory space as it needs to store redundant in-edges and lots of intermediate data.
\\
\subsection{Quality of Graph Partitioning }
\label{subsub:graphPar}
\begin{figure}[h]
	\centering	
	\subfloat[Agent Rate]{
		\includegraphics[scale=0.26]{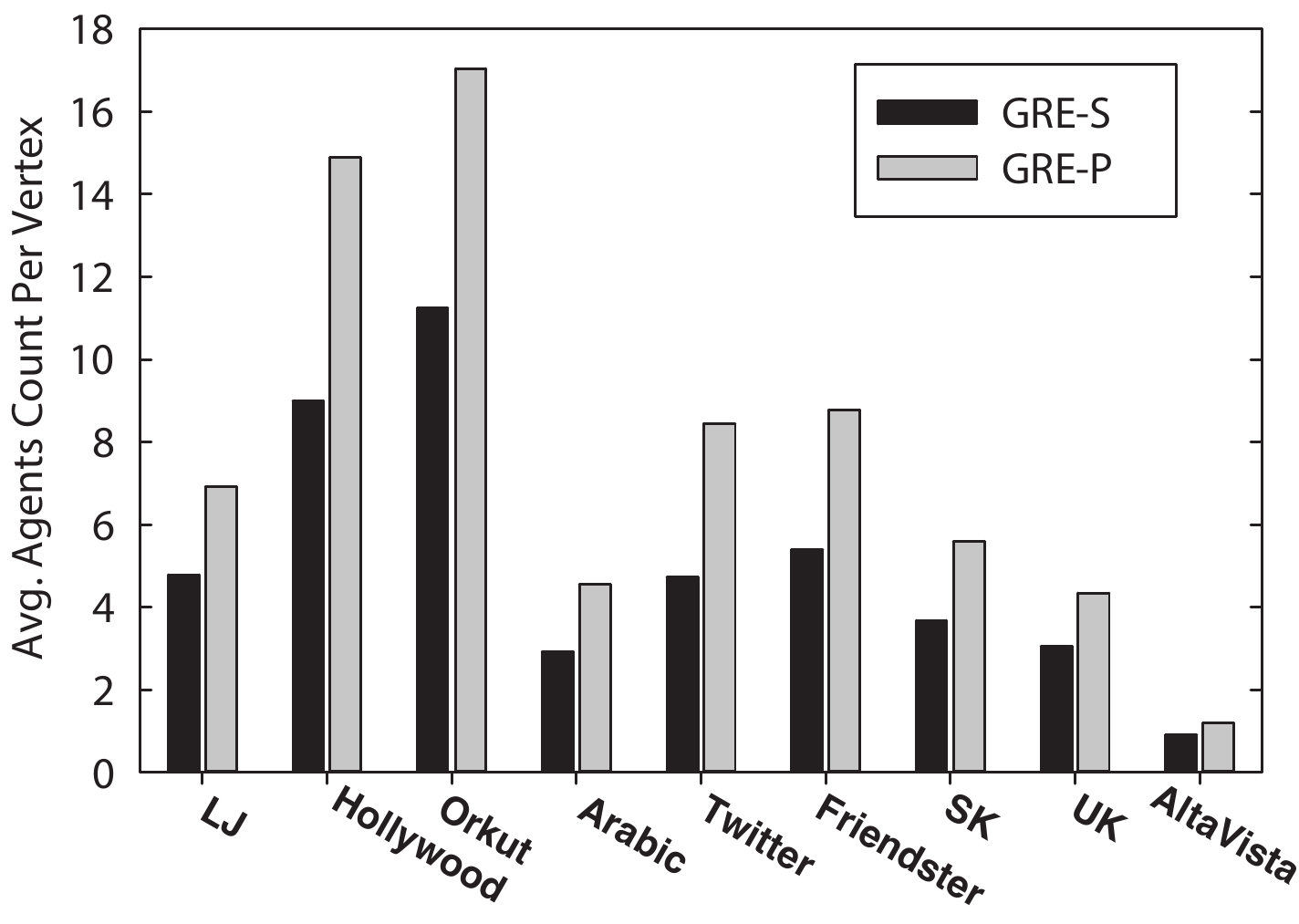}\
		\label{subfig:graphs-par}
	}
	\subfloat[Equivalent Edge-cut Rate]{
		\includegraphics[scale=0.26]{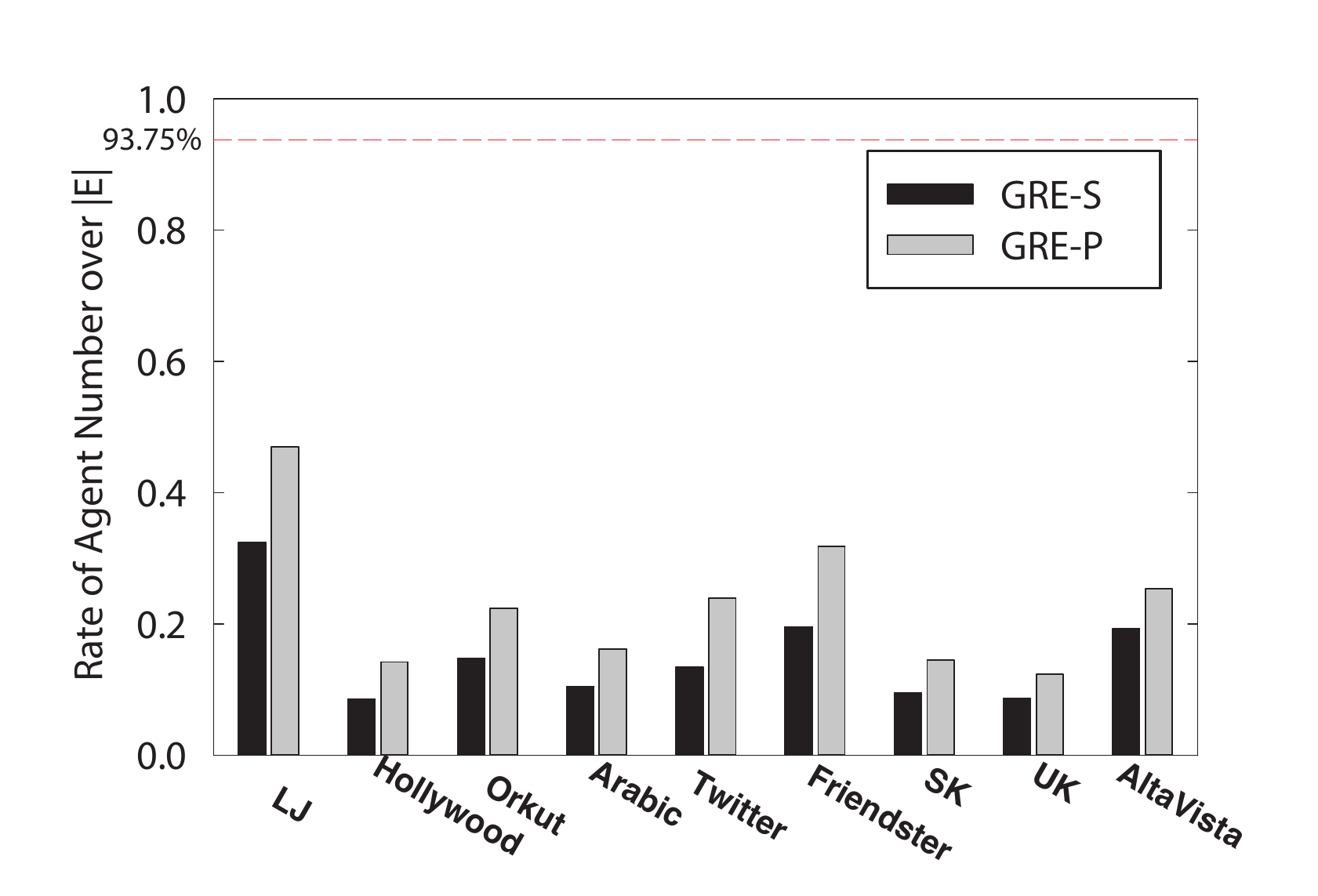}
		\label{subfig:graphs-par-rate}
	}
\caption{Agent-Graph Partitioning of real graphs, with 16 machines. The red dashed line in (b) is edge-cut rate of random hash method.}
\label{fig:par-results}
\end{figure}

The performance advantage of {\tt GRE} is also reflected by the quality of graph partitioning. We first investigate agent-graph partitioning on a broad set of real graphs summarized in Table.\ref{datasets}. Fig. \ref{subfig:graphs-par} shows the average count of agents per vertex, which is translated into {\em equivalent edge-cut rates} in Fig.\ref{subfig:graphs-par-rate} by dividing average vertex degree. As shown in Fig.\ref{subfig:graphs-par-rate}, compared to the traditional random vertex sharding by hashing (red dashed line), agent-graph in both {\tt GRE-P} and {\tt GRE-S} fundamentally reduces cutting edges by  50\%$\sim$91\%.
\vspace{-0.20cm}
\vspace{-0.20cm}

Now, we evaluate scalability of {\tt agent-graph} in terms of the number of partitions, with a comparison to {\tt PowerGraph}'s {\tt vertex-cut}. The partitioning quality metrics {\em cut-factor} is computed according to communication measure: for {\tt agent -graph} the {\em cut-factor} is number of agents (both scatters and combiners) per vertex, while for {\tt vertex-cut} it is $2*(\#Replicas-|V|)$/$|V|$. Higher {\em cut-factor} implies more communication.
\begin{figure}[h]
	\centering	
	\subfloat[Partition Measure]{
		\includegraphics[scale=0.26]{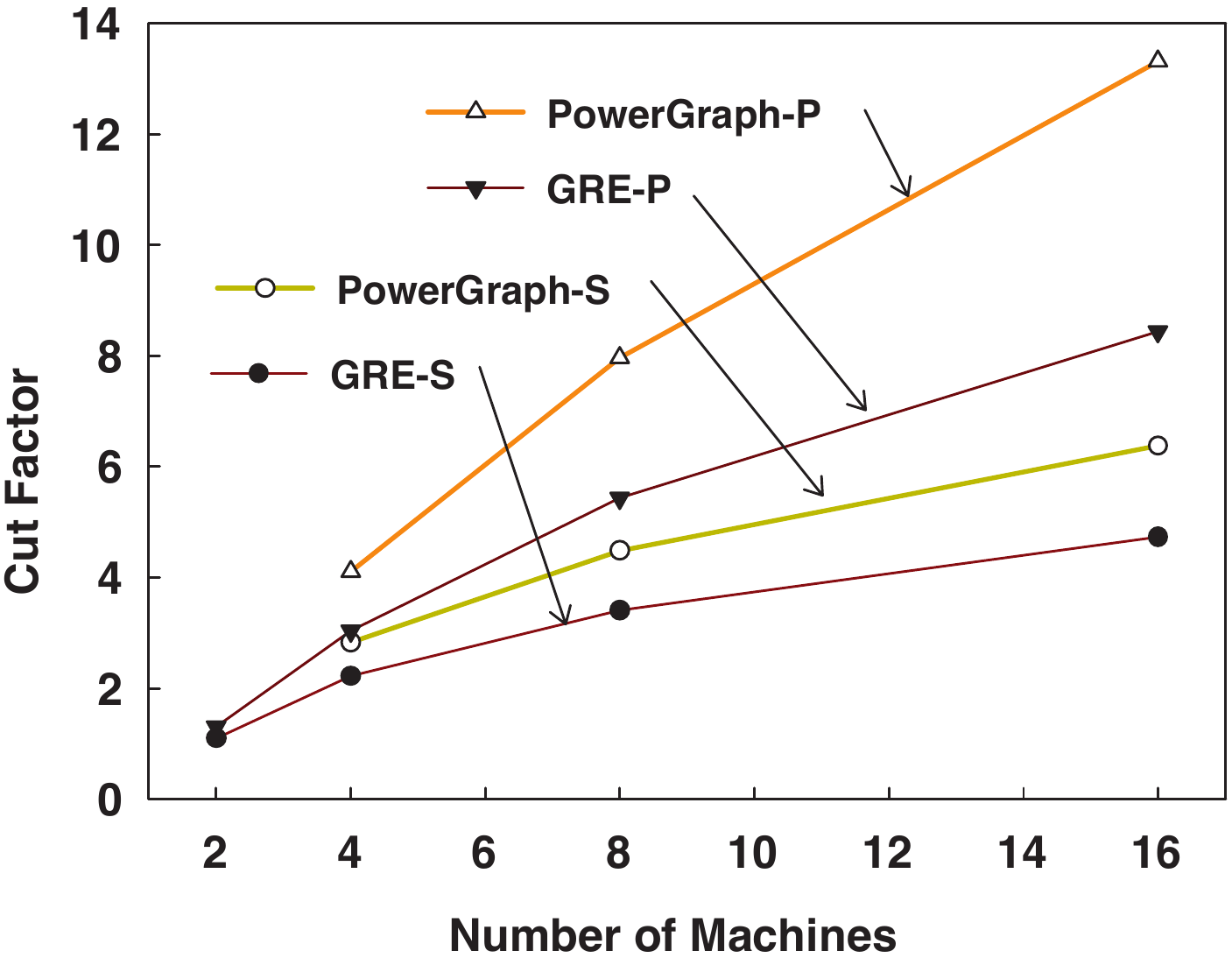}\
		\label{subfig:twitter-par}
	}
	\hspace{0.25cm}
	\subfloat[Agent Skew]{
		\includegraphics[scale=0.26]{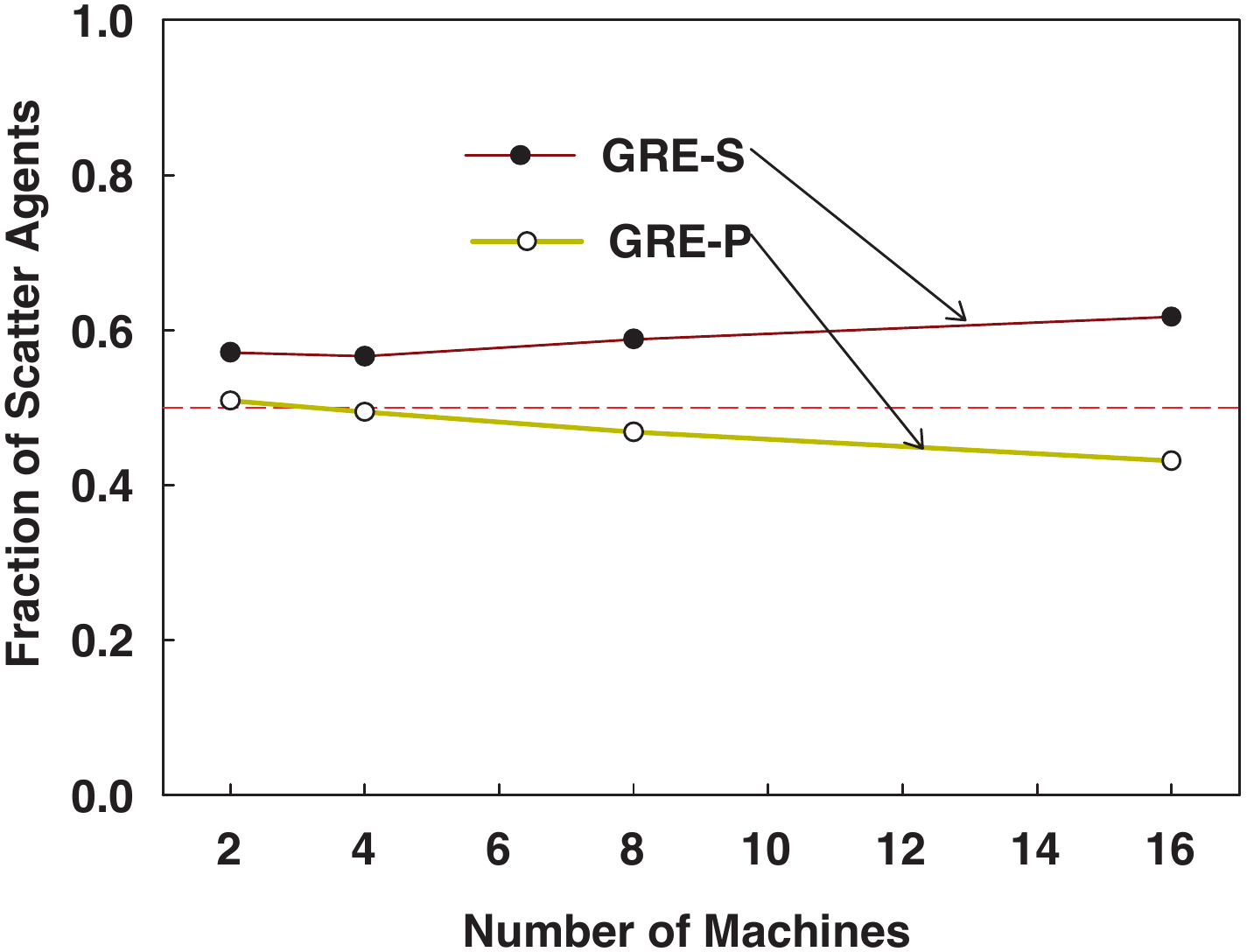}
		\label{subfig:twitter-par-skew}
	}
\caption{Agent-Graph Partitioning Results of Twitter.}
\label{fig:twitter-par}
\vspace{-0.25cm}
\end{figure}
\begin{figure}[h]
	\centering	
	\subfloat[Partition Measure]{
		\includegraphics[scale=0.26]{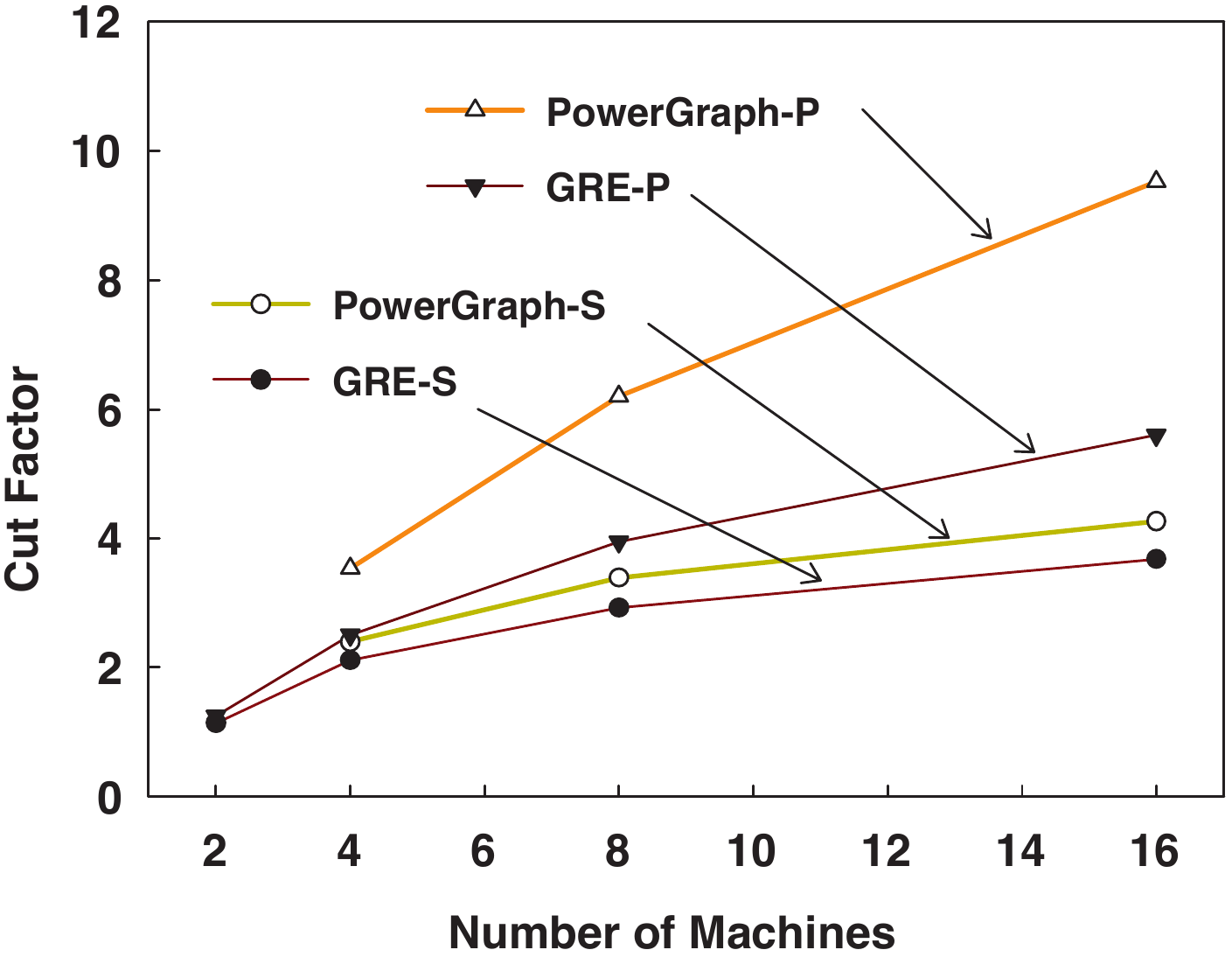}\
		\label{subfig:sk-par}
	}
	\hspace{0.25cm}
	\subfloat[Agent Skew]{
		\includegraphics[scale=0.26]{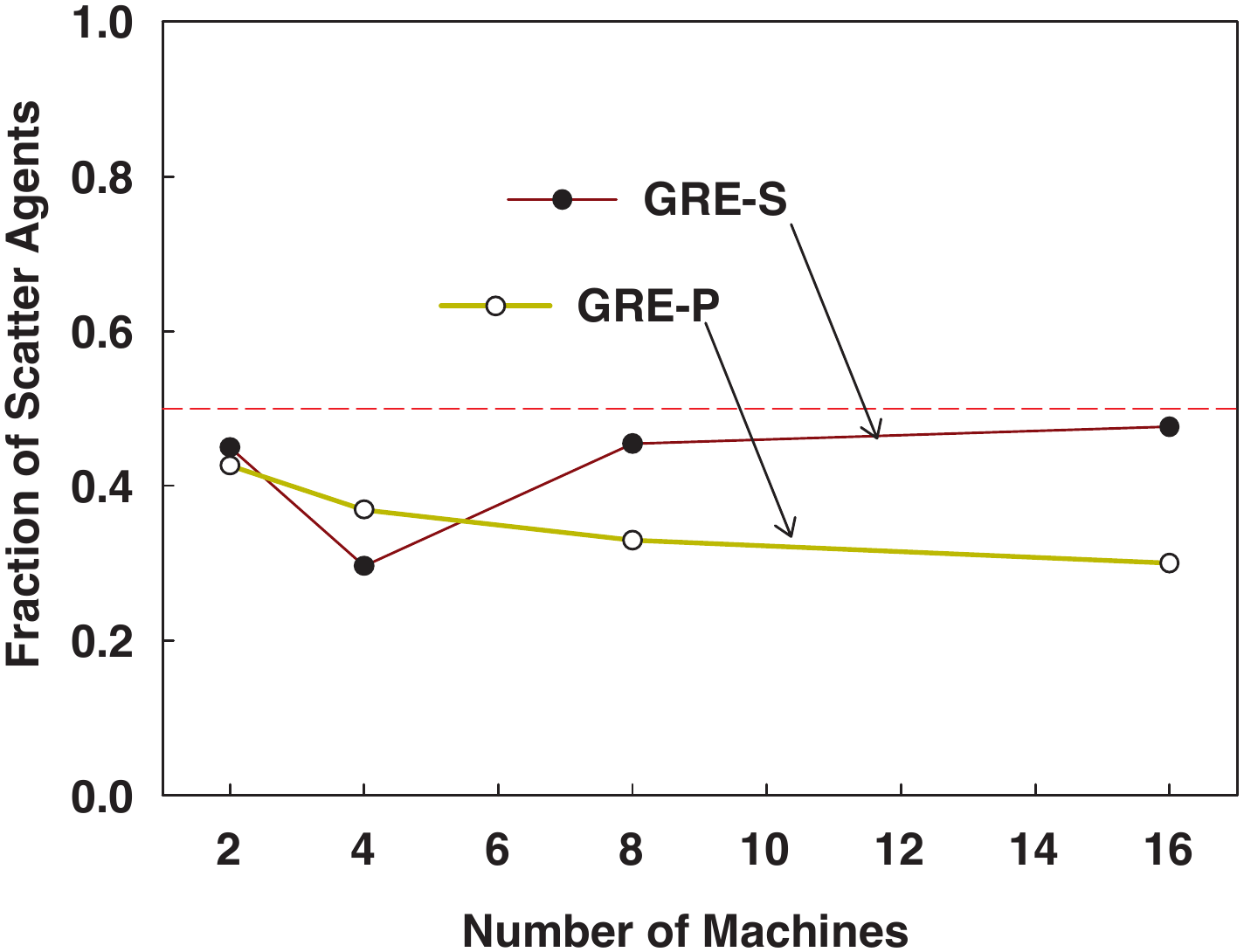}
		\label{subfig:sk-par-skew}
	}
\caption{Agent-Graph Partitioning Results of SK-2005.}
\label{fig:sk-par}
\vspace{-0.30cm}
\end{figure}

Without loss of generality, we choose Twitter social network and SK-2005 web graph for detailed analysis. In real world, social network and web graph are two representative types of graphs. Generally, social networks have comparatively balanced out- and in-degree distribution, while web graphs are typically fan-in. Results of 2$\sim$16 partitions are given in Fig.~\ref{fig:twitter-par} and Fig.~\ref{fig:sk-par}. For both Twitter and SK-2005, {\tt GRE-S} performs best, followed by {\tt PowerGraph-S}, {\tt GRE-P} and {\tt PowerGraph-P} in order. Except for {\tt PowerGraph-P}, all partitioning methods show good scalability over increasing machines (partitions).

To investigate why {\tt GRE-S/P} have better partitioning results than counterparts of {\tt PowerGraph-S/P}, we analyze the percentage distribution of two agent types, i.e. scatter and combiner. As shown in Fig.~\ref{subfig:twitter-par-skew} and Fig.~\ref{subfig:sk-par-skew}, for both {\tt GRE-S} and {\tt GRE-P}, rates of scatters to combiners have an obvious skew. As explained in section \ref{sub:agent-graph-model}, {\tt PowerGraph}'s data model fails to realize this phenomena, while {\tt agent-graph} model can differentiate. Thus, with respect to communication measure, {\tt GRE} has an advantage over {\tt PowerGraph}.

\section{Related Work}
{\tt GRE} adopts the well-known {\em vertex-centric} programming model~\cite{pregel}. Essentially, it is reminiscent of the classic actor model~\cite{actor}. Previously, the most representative vertex-centric abstractions are Pregel~\cite{pregel} and GraphLab~\cite{graphlab}, whose comparison with {\tt GRE} was summarized in Table.~\ref{tab_comp_all}. Here we describe how {\tt GRE} evolves.
\begin{desclist}
\item[Computation model.] Pregel is the first bulk synchronous distributed message passing system. It has been widely cloned in Giraph\cite{giraph}, GPS\cite{gps}, GoldenOrb\cite{goldenorb}, Trinity\cite{trinity} and Mizan\cite{mizan}. Besides, frameworks extending Hadoop (MapReduce) with in-memory iterative execution, such like Spark\cite{spark}, Twister\cite{twister} and HaLoop\cite{haloop}, also adopt a Pregel way to do graph analysis. Meanwhile, GraphLab uses distributed shared memory model and supports both synchronous and asynchronous vertex computation. Vertex computation in both Pregel and GraphLab internally follows the common {\tt GAS} (Gather-Apply-Scatter)\cite{powergraph} pattern. Besides, PowerGraph\cite{powergraph} adopts a phased {\tt GAS}, and can emulate both GraphLab and Pregel. However, {\tt GAS} model handles each edge in a two-sided way that requires two vertices' involvement, leading to amounts of intermediate data storage and extra operations. To address this problem, message {\tt combiner} in Pregel and {\tt delta-caching} in PowerGraph are proposed as complement to basic {\tt GAS}. Instead of {\tt GAS}, {\tt GRE} proposes a new {\tt Scatter-Combine} model, which explicitly transforms {\tt GAS}'s two-sided edge operation into one-sided active message. In the worst case, active message can degrade to message passing of Pregel. 
\item[Distributed graph model.] {\tt GRE}'s {\tt Agent-Graph} model is derived from optimizing message transmission on directed graphs. Previously, Pregel has introduced {\em combiner} for combining messages to the same destination vertex. {\tt GRE} further introduces {\em scatter} to reduce messages from the same source vertex. Motivated by {\em ghost} in PBGL\cite{pbgl} and GraphLab, we finally develop ideas of message {\em agent} into a distributed directed graph model. Note that {\tt GPS}\cite{gps}, an optimized implementation of {\tt Pregel}, supports Large Adjacency-List Partitioning where the {\em subvertex} is similar to {\tt scatter} on reducing messages but not well-defined for vertex or edge computation. The closest match to {\tt Agent-Graph} is PowerGraph's {\tt vertex-cut} which however is used only in undirected graphs and coupled with different computation and data consistency models.
\end{desclist}

Besides the vertex-centric model, generalized {\tt SpMV}  (Sparse Matrix-Vector) computation is another popular graph-parallel abstraction, which is used by Pegasus \cite{pegasus} and Knowledge Discovery Toolbox \cite{kdt}. Note that since {\tt SpMV} computation is naturally bulk synchronous and factorized over edges, their applications can be described in {\tt GRE}. However, unlike {\tt GRE}, the matrix approach is not suitable for handling abundant vertex/edge metadata.

For shared memory environment, there are also numerous graph-parallel frameworks. Ligra~\cite{ligra} proposes an abstraction of edgeMap and vertexMap which is simple but efficient to describe traversal-based algorithms. GraphChi~\cite{graphChi} uses a sliding window to process large graphs from disks in just a PC. X-Stream \cite{x-stream} proposes a novel edge-centric scatter-gather programming abstraction for both in-memory and out-of-core graph topology, which essentially, like {\tt GRE} and {\tt PowerGraph}, leverages vertex factorization over edges. {\tt GRE}'s computation on local machine is highly optimized for massive edge-grained parallelism, based on technologies such as vLock~\cite{vlock} fine-grained data synchronization and FastForward~\cite{fastforward} thread-level communication.

\section{Conclusions}
\label{conclusion}
Emerging {\em graph-parallel} applications have drawn great interest for its importance and difficulty. We identify that the performance-related difficulty lies on two aspects, i.e. irregular parallelism expressing and graph partitioning. We propose {\tt GRE} to address these two problems from both computation model and distributed graph model. First, the {\tt Scatter-Combine} model retains the classic vertex-centric programing model, and realizes the irregular parallelism by factorizing vertex computation into a series of active messages in parallel. Second, along the idea of vertex factorization, we develop distributed {\tt Agent-Graph} model that can be constructed by a vertex-cut way like in PowerGraph. Compared to traditional edge-cut partitioning methods or even PowerGraph's vertex-cut approach, {\tt Agent-Graph} significantly reduces communication. Finally, we develop an efficient runtime system implementation for {\tt GRE}'s abstractions and experimentally evaluate it with three applications on both real-world and synthetic graphs. Experiments on our 16-node cluster system demonstrate {\tt GRE}'s advantage on performance and scalability over other counterpart systems.

\balance



{\scriptsize
\begin{thebibliography}{50}
\vspace{0.25cm}
\bibitem{law}
http://law.di.unimi.it

\bibitem{webScale}
http://www.worldwidewebsize.com

\bibitem{facebookScale}
http://www.facebook.com/press/info.php?statistics

\bibitem{graph500}
http://www.graph500.org

\bibitem{altavista}
http://webscope.sandbox.yahoo.com

\bibitem{giraph}
http://giraph.apache.org

\bibitem{goldenorb}
http://goldenorbos.org/

\bibitem{sk}
P. Boldi, B. Codenotti, M. Santini and S. Vigna. UbiCrawler: A Scalable Fully Distributed Web Crawler. Software: Practice \& Experience. 32(8): 711--726, 2004.

\bibitem{lj}
F. Chierichetti, R. Kumar, S. Lattanzi, M. Mitzenmacher, A. Panconesi and P. Raghavan. On compressing social networks. In KDD, 2009.

\bibitem{orkut}
A. Mislove, M. Marcon, K. Gummadi, P. Druschel and B. Bhattacharjee. Measurement and Analysis of Online Social Networks. In IMC, 2007.

\bibitem{hollywood}
P. Boldi, M. Rosa, M. Santini and S. Vigna. Layered Label Propagation: A MultiResolution Coordinate-Free Ordering for Compressing Social Networks. In WWW, 2011.

\bibitem{uk}
P. Boldi, B. Codenotti, M. Santini and S. Vigna. A Large Time-Aware Graph. SIGIR Forum, 42(2):33--38, 2008.

\bibitem{twitter}
H. Kwak, C. Lee, H. Park and S. Moon. What is Twitter, a Social Network or a News Media?. In WWW, 2010.

\bibitem{streamingPar}
I. Santon and G. Kilot. Streaming Graph Partitioning for Large Distributed Graphs. In KDD, 2012.

\bibitem{pregel}
G. Malewicz, M. Austern, A. Bik, J. Dehnert, I. Horn, N. Leiser and G. Czajkowski. Pregel: a system for large-scale graph processing. In SIGMOD, 2010.

\bibitem{graphlab}
Y. Low, J. Gonzalez, A. Kyrola, D. Bickson, C. Guestrin and M. Hellerstein. Distributed GraphLab: A Framework for Machine Learning and Data Mining in the Cloud. In PVLDB, 2012.

\bibitem{powergraph}
J. Gonzalez, Y. Low, H. Gu, D. Bickson, C. Guestrin. PowerGraph: Distributed Graph-Parallel Computation on Natual Graphs. In OSDI, 2012.

\bibitem{spark}
M. Zaharia, M. Chowdhury, S. Shenker and I. Stoica. Spark: Cluster computing with working sets. In HotCloud, 2010.

\bibitem{hardPar1}
A. Abourjeili and G. Karypis.  Multilevel algorithms for partitioning power-law graphs. In IPDPS, 2006.

\bibitem{hardPar2}
J. Leskovec, K. Lang, A. Dasgupta and W. Mahoney. Community structure in large networks: Natural cluster sizes and the absence of large well-defined clusters. Internet Mathematics 61 (2008), 29--123.

\bibitem{pbgl}
D. Gregor and A. Lumsdaine. The parallel BGL: A generic library for distributed graph computations. In POOSC, 2005.

\bibitem{cos}
M. Stonebraker, D.J. Abadi, A. Batkin, X. Chen, M. Cherniack, M. Ferreira, E. Lau, A. Lin, S. Madden, E. O'Neil, P. O'Neil, A. Rasin, N. Tran, and S. Zdonik. C-store: a column-oriented DBMS. In VLDB, 2005.

\bibitem{signal_collect}
S. Philips, B. Abraham and C. William. Signal/Collect : Graph Algorithms for the Web. In ISWC, 2010.

\bibitem{bsp}
L. Valliant. A bridging model for parallel for parallel computation. Comm. ACM 33(8), 1990, 103-111.

\bibitem{agentPar}
J. Yan, G.M. Tan, and N.H Sun. Agent-Graph: A Distributed Data Model for Scale-free Graphs, Technical Report, ICT, 2013.

\bibitem{vlock}
J. Yan, G. Tan, X. Zhang, E. Yao and N. Sun. vLock: Lock Virtualization Mechanism for Exploiting Fine-grained Parallelism in Graph Traversal Algorithms. In CGO, 2013.

\bibitem{challenges}
A. Lumsdaine, D. Gregor, B. Hendrickson, and J. W. Berry. Challenges in parallel graph processing. Parallel Processing Letters, 7(1): 5--20, 2007.

\bibitem{sssp}
D.~P. Bertsekas, F.~Guerriero, and R.~Musmanno. Parallel asynchronous label correcting methods for shortest paths. Journal of Optimization Theory and Applications, 89, 1996.

\bibitem{pagerank}
S.~Brin and L.~Page. The anatomy of a large-scale hypertextual web search engine. In Computer Networks and ISDN systems, pages 107--117. Elsevier Science Publishers, 1998.

\bibitem{trinity}
B. Shao, H. Wang and Y. Li. Trinity: A Distributed Graph Engine on a Memory Cloud. In SIGMOD, 2013.

\bibitem{csr}
Y. Saad. SPARSKIT: A basic toolkit for sparse matrix computations. Technical Report, NASA Ames Research Center, 1990.

\bibitem{bellman1956routing}
R. Bellman. On a routing problem. Quarterly of Applied Mathematics, 16(1), 1958, 87--90.

\bibitem{introAlg}
T.H. Cormen, C.E. Leiserson, R.L. Rivest and C. Stein. Introduction to Algorithms. MIT Press, 2009.

\bibitem{friendster}
J. Yang and J. Leskovec. Defining and Evaluating Network Communities based on Ground-truth. In ICDM, 2012.

\bibitem{actor}
C. Hewitt, P. Bishop and R. Steiger. A Universal Modular Actor Formalism for Artificial Intelligence. In IJCAI, 1973.

\bibitem{active_messages}
T. Von Eicken, D. Culler, S. Goldstein, and K. Schauser. Active messages: a mechanism for integrated communication and computation. In ISCA, 1992.

\bibitem{pegasus}
U. Kang, C. Tsourakakis, A. Appel, and C. Faloutsos. PEGASUS: mining peta-scale gaphs. Knowl. Inf. Syst., 27(2), 2011.

\bibitem{kdt}
A. Luguwski, D. Albert, A. Buluc, J. Gilbert, S. Reinhart, Y. Teng and A. Waranis. A flexible open-source toolbox for scalable complex graph analysis. In SDM, 2012.

\bibitem{bipartite-match}
T. Anderson, S. Owicki, J. Saxe and C. Tracker. High-speed Switch Schedule for Local-Area Networks. ACM Tran. Comp. Syst. 11(4), 1993, 319-352.

\bibitem{bc}
L. Freeman. A set of measures of centrality based upon betweenness. Sociometry. 1997.

\bibitem{lbp}
C. Bishop. Pattern Recognition and Machine Learning. 2nd edition. Springer. 2007.

\bibitem{x-stream}
A. Roy, I. Mihailovic and W. Zwaenepoel. X-Stream: Edge-Centric Graph Processing using Streaming Partitions, in SOSP, 2013.

\bibitem{twister}
J. Ekanayake, H. Li, B. Zhang,  T. Gunarathne, S. Bae, J. Qiu, and G. Fox. Twister: A runtime for iterative MapReduce. In HPDC, 2010.

\bibitem{gps}
S. Salihoglu and J. Widom. GPS: A Graph Processing System. Tech. Report, Standford University, 2012.

\bibitem{mizan}
Z. Khayyat, K. Awara, A. Alonazi, H. Jamjoom, D. Williams, and P. Kalnis. Mizan: a system for dynamic load balancing in large-scale graph processing. In EuroSys, 2013.

\bibitem{mapreduce}
J. Dean and S. Ghemawat. MapReduce: Simplified Data Processing on Large Clusters. In OSDI, 2004.

\bibitem{haloop}
Y. Bu, B. Howe, M. Balazinska and M. Ernst. HaLoop: Efficient iterative data processing on large clusters. In VLDB, 2010.


\bibitem{ligra}
J. Shun and G. Belloch. Ligra: A Lightweight Graph Processing Framework for Shared Memory. In PPoPP, 2013.

\bibitem{fastforward}
J. Giacomoni, T. Moseley and M. Vachharajani. FastForward for efficient pipeline parallelism: a cache-optimized concurrent lock-free queue. In PPoPP, 2008.

\bibitem{optimize_pregel}
S. Salihoglu and J. Widom. Optimizing Graph Algorithms on Pregel-like Systems. Technical Report. Stanford InfoLab. 2013.

\bibitem{graphChi}
A. Kyrola, G. Blelloch, and C. Guestrin. GraphChi: Large-Scale Graph computation on Just a PC. In OSDI, 2012.

\end{thebibliography}
}

\end{document}